\documentclass[11pt]{article}

\usepackage{amsmath,amsfonts,amssymb}

\numberwithin{equation}{section}

\newcommand{\be}{\begin{equation}}
\newcommand{\ee}{\end{equation}}
\newcommand{\ba}{\begin{eqnarray}}
\newcommand{\ea}{\end{eqnarray}}

\newcommand{\ns}{\normalsize}

\newcommand{\ax}{\alpha}
\newcommand{\bx}{\beta}

\newcommand{\dx}{\delta}

\newcommand{\gx}{\gamma}
\newcommand{\ab}{\bar\alpha}
\newcommand{\bb}{\bar\beta}
\newcommand{\cb}{\bar\gamma}

\newcommand{\Ox}{\Omega}

\newcommand{\Gx}{\Gamma}

\newcommand{\cN}{\mathcal{N}}
\newcommand{\cG}{\mathcal{G}}

\newcommand{\cF}{\mathcal{F}}
\newcommand{\cH}{\mathcal{H}}

\newcommand{\cQ}{\mathcal{Q}}
\newcommand{\cM}{\mathcal M}
\newcommand{\W}{\mathcal W}
\newcommand{\V}{\mathcal V}

\newcommand{\wg}{\wedge}

\newcommand{\nn}{\nonumber}
\def\theequation{\arabic{section}.\arabic{equation}}

\newcommand{\gK}{h^{(1)}}
\newcommand{\gKi}{h_{(1)}}
\newcommand{\gcs}{h^{(2)}}
\newcommand{\gcsi}{h_{(2)}}
\newcommand{\KK}{K^{(1)}}
\newcommand{\Kcs}{K^{(2)}}

\def\a{\alpha}
\def\b{\beta}
\def\g{\gamma}
\def\c{\chi}
\def\d{\delta}
\def\e{\epsilon}

\def\f{\phi}

\def\z{\psi}
\def\k{\kappa}
\def\l{\lambda}
\def\m{\mu}
\def\n{\nu}
\def\o{\omega}

\def\r{\rho}
\def\s{\sigma}

\def\x{\xi}
\def\z{\zeta}

\def\F{\Phi}
\def\G{\Gamma}

\def\O{\Omega}

\begin{document}


\begin{titlepage}

\title{
   \hfill{\ns YITP-04-45\\}
   \hfill{\ns hep-th/0408121\\}
   \vskip 2cm
   {\Large\bf Heterotic on Half-flat}\\[0.5cm]}
   \setcounter{footnote}{0}
\author{
{\ns\large
    Sebastien Gurrieri$^1$\footnote{email: gurrieri@yukawa.kyoto-u.ac.jp}}~,
\setcounter{footnote}{3}
{\ns\large Andr\'e Lukas$^2$\footnote{email: A.Lukas@sussex.ac.uk}}~
  {\ns and Andrei Micu$^2$\footnote{email: A.Micu@sussex.ac.uk}
   $^{,}$\footnote{On leave from IFIN-HH Bucharest.}}
  \\[0.5cm]
   {\it\ns $^1$Yukawa Institute for Theoretical Physics, Kyoto University}\\
   {\ns Kyoto 606-8502, Japan} \\[0.5cm]
   {\it\ns $^2$Department of Physics and Astronomy, University of Sussex}\\
   {\ns Brighton BN1 9QJ, UK}}
\date{}

\maketitle

\begin{abstract}\noindent
We study the effective action of the heterotic string compactified on
particular half-flat manifolds which arise in the context of mirror
symmetry with NS-NS flux. We explicitly derive the superpotential and
K\"ahler potential at lowest order in $\a '$ by a reduction of the
bosonic action. The superpotential contains new terms depending on
the K\"ahler moduli which originate from the intrinsic geometrical
flux of the half-flat manifolds. A generalized Gukov formula, valid
for all manifolds with ${\rm SU}(3)$ structure, is derived from the
gravitino mass term. For the half-flat manifolds it leads to a
superpotential in agreement with our explicit bosonic calculation.
We also discuss the inclusion of gauge fields.
\end{abstract}

\vfill

August 2004

\thispagestyle{empty}

\end{titlepage}


\section{Introduction}

The stabilization of moduli remains one of the central problems when
trying to relate string theory to low-energy particle
physics. Recently, flux compactifications were intensively studied as
a method to tackle this problem, mostly in the context of type II
strings or
M-theory~\cite{PS}--\cite{BC2}. The analysis is particularly straightforward
within the context of type IIB strings on Calabi-Yau spaces where a
combination of NS-NS and RR flux can be used to fix all complex
structure moduli as well as the axion-dilaton~\cite{GKP}. If all
moduli are successfully stabilized in such models~\cite{KKLT} the
radius of the internal space is usually not much larger than the
string scale. This excludes very large additional dimensions and a low
string scale and means that low-energy supersymmetry remains as the
only known option to stabilize the electroweak scale. Explicit
examples for type II brane models can be found, for example, in
Refs.~\cite{Aldazabal:1999tw}--\cite{Blumenhagen:2002gw}. The
construction of phenomenologically attractive {\em supersymmetric}
type II brane models has so far proven difficult, however, see
Ref.~\cite{MS}.

\vspace{0.4cm}

The situation is somewhat reversed in the context of heterotic string
models. It has been known for a long time that supersymmetric models
with broadly the right phenomenological properties can be obtained
easily and in large numbers~\cite{CHSW,GSW}. NS-NS flux in heterotic
compactifications has been introduced some time
ago~\cite{DRSW,RW,WW,AS,CH} and there are also a number of more recent
discussions~\cite{GKLM,BBHL,BCtin,BD1,HPN} of the subject. However,
discarding the $E_8\times E_8$ or ${\rm SO}(32)$ gauge fields whose
vacuum expectation values are tied to the curvature via the Bianchi
identity, the NS-NS three-form field strength is the only
antisymmetric tensor field in heterotic theories which implies an
apparent lesser degree of flexibility in fixing moduli through flux,
as compared with type II theories. In particular, no even-degree form
field strength is available whose flux could fix the K\"ahler moduli.

\vspace{0.4cm}

In this paper, we are going to address this problem by considering the
heterotic string on particular six-dimensional manifolds with ${\rm
SU}(3)$ structure and non-vanishing (intrinsic) torsion. We will see
that these manifolds encode even-degree flux ``geometrically'' and we
will compute the resulting K\"ahler moduli superpotential explicitly.
The existence of these manifolds is suggested by type II mirror
symmetry with NS-NS flux, as has been argued in Ref.~\cite{GLMW,GM}.
This conjecture was generalized and further evidence was provided for
it in Refs.~\cite{FMT,GMPT}. Explicit non-compact examples
were constructed in Refs.~\cite{BDKT,ABBDKT}. Also, consistency of embedding
such backgrounds in string/ M-theory was discussed in Ref.~\cite{MM}.

In Ref.~\cite{GLMW}, it was proposed that type IIB
(IIA) on a Calabi-Yau three-fold with NS-NS flux is mirror-symmetric
to IIA (IIB) on a particular class of six-dimensional half-flat
manifolds with ${\rm SU}(3)$ structure. In the following, we will
refer to these manifolds as half-flat mirror manifolds.  Under the
mirror map, the original odd-degree NS-NS flux, which generates a
superpotential for the complex structure moduli, is mapped to
even-degree geometrical flux of the half-flat mirror manifolds, which
generates a superpotential for the K\"ahler moduli. In this paper, we
are not interested in type II mirror symmetry by itself but merely as
a means of ``defining'' the half-flat mirror manifolds. Our goal is to
consider the heterotic string on the so-defined manifolds with torsion.

\vspace{0.4cm}

The heterotic string on non-K\"ahler manifolds was already discussed
in a number of papers~\cite{CCDLMZ,BBDG,BBDP,CCDL,BBDGS,BD} and the
supersymmetric
solutions were classified in terms of the five torsion classes of
manifolds with ${\rm SU}(3)$ structure. However, the lack of knowledge
of internal properties (such as the moduli space) of general manifolds
with ${\rm SU}(3)$ structure makes it difficult to derive the
effective action for such theories and, so far, general properties of
the superpotentials have been discussed~\cite{BBDG,BBDP,CCDL}.

In comparison, we can see a number of advantages in our
approach. First of all, type II mirror symmetry strongly suggests the
existence of the half-flat mirror manifolds and it imposes very strong
constraints on them. In fact, mirror symmetry provides us with a
concrete set of relations describing half-flat mirror manifolds, which
allows the calculation of much of the low-energy effective action.
Their mirror symmetry origin implies that a half-flat mirror manifold
should exist for each Calabi-Yau three-fold with a mirror and for each
set of NS flux parameters. Hence, we are dealing with a large class of
manifolds which is closely linked to Calabi-Yau three-folds.  This
will hopefully lead to models preserving many of the attractive
features of heterotic Calabi-Yau compactifications while at the same time
enhancing the flexibility for moduli stabilization through flux.

\vspace{0.4cm}

In this paper, we will mainly focus on zeroth order in $\a '$,
that is, on the gravitational sector of the heterotic string and we
will discuss only some aspects of including gauge fields. The full
gauge field sector will be included in a forthcoming
publication~\cite{GLM}. Our main aim is to derive the effective
four-dimensional $N=1$ supergravity for the heterotic string on
half-flat mirror manifolds to this order in $\a '$. In particular, we
will compute the superpotential which will be done in two largely
independent ways, namely from the bosonic action and the gravitino
mass term. We also obtain a general Gukov-type formula for the
superpotential which we expect to hold for all heterotic
compactifications on manifolds with ${\rm SU}(3)$ structure and to
first order in $\a '$.

The outline of the paper is as follows. In the next section, we
present a brief review of the 10-dimensional action of the
heterotic string and of the half-flat mirror manifolds on which we
are going to carry out the dimensional reduction. Section 3
reduces the bosonic part of the action on half-flat mirror
manifolds at zeroth order in $\a '$, at first without and
eventually including NS-NS flux. In section 4, we present an
alternative derivation of the superpotential from fermionic
terms in the action. Based on the gravitino mass term, we
first derive a Gukov-type formula for the heterotic string on
${\rm SU}(3)$ structure manifolds and then show that, specialized
to half-flat mirror manifolds, it reproduces the previous result
for the superpotential. Section 5 discusses some steps necessary
to include gauge fields and we conclude in Section 6. Two
appendices present some relevant results in special geometry and
the calculation of the potential from the superpotential in the
general case.


\section{Review of background material}

In this section, we present some background material in order to set
up our notation and conventions. Firstly, we review the 10-dimensional
effective action of the heterotic string~\cite{GSW} which is the
action we would like compactify to four dimensions. Then we describe
the half-flat mirror manifolds~\cite{GLMW} on which we are
going to carry out the dimensional reduction.

\subsection{Ten-dimensional effective action for the heterotic string}

The 10-dimensional effective action for the heterotic string is
given, to leading order in $\alpha '$, by 10-dimensional $N=1$
supergravity coupled to 10-dimensional super-Yang-Mills theory
with gauge group $E_8\times E_8$ or ${\rm SO}(32)$. In this paper,
we will focus on the $E_8\times E_8$ case for definiteness but most
of our considerations will directly apply to the ${\rm SO}(32)$
case as well. Ten-dimensional coordinates are denoted by $(x^M)$,
labeled by curved indices $M,N,\dots =0,\dots ,9$.

The 10-dimensional $N=1$ supergravity multiplet consists of the
metric $\hat{g}_{MN}$, the dilaton $\hat{\f}$, the NS-NS two-form
$\hat{B}_{MN}$ and their fermionic partners, the gravitino
$\hat{\Psi}_M$ and the dilatino $\hat{\l}$, both 10-dimensional
Majorana-Weyl spinors which we take to be of positive chirality. Here
and in the following a hat denotes a 10-dimensional quantity. To
lowest (zeroth) order in the $\a '$ expansion the bosonic part of the
effective action is given by~\cite{GSW}
\begin{equation}
  \label{SB}
  S_{0,{\rm bosonic}} = -\frac{1}{2\k_{10}^2} \int_{M_{10}}
  e^{-2 \hat \phi} \left[ \hat R \star \mathbf{1}
    - 4 d \hat \phi \wedge\star  d \hat \phi
    + \frac12 \hat H \wedge \star \hat H \right] \, ,
\end{equation}
where $\hat{H}= d \hat{B}$ is the three-form field strength of $\hat{B}$
and $\hat{R}$ is the Riemann curvature scalar. We will later find it
useful to consider some of the fermionic terms. To zeroth order in
$\a'$ they read
\begin{eqnarray}
  \label{SF}
  S_{0,{\rm fermionic}} & = & - \frac{1}{2} \int_{M_{10}} d^{10} x
  \sqrt{-\hat{g}} \, e^{-2 \hat \phi}  
  \Bigg\{ \overline{\hat\Psi}_M\G^{MNP} D_N\hat \Psi_P \\
  & & \hspace{2cm} - \frac{1}{24}  \left(
  \overline{\hat\Psi}_M \G^{MNPQR} \hat \Psi_R
  + 6 \, e^{-\hat \phi} \, \overline{\hat\Psi}^N \Gx^P \hat \Psi^Q
  \right) \hat{H}_{NPQ}+\; \dots \Bigg\} \, , \nn
\end{eqnarray}
where the dots stand for additional four-fermion terms and terms which
involve the dilatino.
Here, $\G_M$ are the 10-dimensional gamma matrices which are taken to
be real, conjugation is defined as $\bar{\psi}=\psi^\dagger\G_0$ for a
spinor $\psi$ and multi-indexed $\G$ symbols denote anti-symmetrized
products of gamma matrices with unit norm, as usual. For convenience
we have chosen the overall dilaton factor to be the same as in the
bosonic part of the action by appropriately rescaling the gravitino
field.

\vspace{0.4cm}

The 10-dimensional Yang-Mills multiplet consists of the gauge
field $\hat{A}_M$, with field strength $\hat{F}_{MN}$, and its
superpartner, the gaugino, both in the adjoint $E_8\times E_8$
(or ${\rm SO}(32)$).  The kinetic terms for these fields along
with $\hat{R}^2$ terms and additional four-fermion terms involving
the gauginos arise at order $\a '$. The bosonic among those terms
are given by
\begin{equation}
 S_{1,{\rm bosonic}} = -\frac{\a '}{16\k_{10}^2} \int_{M_{10}}d^{10}x
 \sqrt{-\hat{g}} e^{-\hat{\phi}}
 \left\{{\rm Tr}( \hat{F}^2) - {\rm tr}( \tilde{R}^2) \right\}
\end{equation}
where ${\rm tr}(\tilde{R}^2)$ really stands for the Gauss-Bonnet
combination. The curvature two-form $\tilde{R}$ is computed in terms
of the modified connection
\begin{equation}
 \tilde{\o}_{IJ}{}^K = \o_{MN}{}^P+\frac{1}{2}\hat{H}_{MN}{}^P\; ,
 \label{ot}
\end{equation}
where $\o$ is the Levi-Civita connection. The other modification to
the action at this order appears in the definition of the field
strength $\hat{H}$ which now becomes
\begin{equation}
 \hat{H} = d\hat{B}+\frac{\a '}{4}\left(\o_{\rm L}-\o_{\rm YM}\right)\; .
 \label{BI}
\end{equation}
Here, $\o_{\rm L}$ and $\o_{\rm YM}$ are the usual Lorentz and
Yang-Mills Chern-Simons three-forms defined by
\begin{eqnarray}
 \o_{\rm L} &=& {\rm tr} \left( \tilde{R} \wedge \tilde{\o} -
   \frac{1}{3} \tilde{\o} \wedge \tilde{\o} \wedge \tilde{\o} \right) \\
 \o_{\rm YM} &=& {\rm Tr} \left( \hat{F} \wedge \hat{A} - \frac{1}{3}
   \hat{A} \wedge \hat{A} \wedge \hat{A} \right)
\end{eqnarray}
The trace ${\rm Tr}$ denotes $1/30$ of the trace in the adjoint for
$E_8\times E_8$ or the trace in the fundamental for ${\rm SO}(32)$, as
usual. These are the only corrections to the action at order
$\a'$. Further
terms appear at order ${\a '}^2$ which, however, will not
concern us here. In fact, throughout most of the paper we will focus
on the leading, zeroth order in $\a '$ for which we present a complete
analysis. In addition, we discuss some aspects related to the
gauge fields.


\subsection{Half-flat manifolds}

We will now briefly describe the particular six-dimensional manifolds
on which we are going to carry out the reduction of the 10-dimensional
heterotic effective action. In general terms, these manifolds arise as
the mirrors of Calabi-Yau manifolds with (a particular type of) NS-NS
flux, as constructed in Ref.~\cite{GLMW}.  Before we get to this
specific definition in terms of mirror symmetry it is useful to review
the main properties of the general manifolds with ${\rm SU}(3)$
structure and their classification in terms of torsion classes
following \cite{CS} and then specialize to the case
of half-flat mirror manifolds.

\vspace{0.4cm}

A six-dimensional manifold is said to have ${\rm SU}(3)$ structure if
it admits a globally defined spinor~\footnote{For definiteness we will
take this spinor to be Weyl, but one can as well work with Majorana
spinors.} which we denote $\eta$. From a physical point of view this
is the most practical definition as this globally defined spinor
ensures that the action obtained by compactifying on such manifolds
preserves some supersymmetry.

The geometric properties of manifolds with ${\rm SU}(3)$ structure are
better described in terms of two invariant forms $J$ and $\Ox$ which
can be defined as bi-linears in the spinor $\eta$ as follows
\begin{equation}
  \label{Oeta}
  \begin{aligned}
    J_{mn} =& -i\eta^\dagger\g_{mn}\eta \ , \\
    \O_{mnp} =& -\frac{i ||\Ox||}{\sqrt 8} \eta^\dagger \g_{mnp} \,
    \eta^* \; \ .
  \end{aligned}
\end{equation}
Here, $\gx_m$, with indices $m,n\dots = 5,\dots , 9$ are
six-dimensional Euclidean gamma matrices which are chosen to be purely
imaginary. As before, multiple indices denote antisymmetrisation.
Note that the normalization of $\Ox$ is different from what can be
found in the literature and was chosen in order to agree with the
usual moduli space conventions. Indeed, it is easy to check using
gamma matrix algebra and Fierz identities that
\begin{equation}
  \Ox_{mnp} \bar \Ox^{mnp} = 3! ||\Ox||^2 \ ,
\end{equation}
provided the spinor $\eta$ satisfies $\eta^\dagger \eta =1$.


Manifolds with ${\rm SU}(3)$ structure can be classified by their
intrinsic torsion and it will be useful to briefly review this. For a
more complete account see, for example, Refs.~\cite{CS}.
It is well known that the ${\rm SU}(3)$ structure
induces a metric on the manifold \cite{Hitchin}. The Levi-Civita
connection associated to this metric violates in general the
structure, but there always exists a connection which we denote
$\nabla^{\rm (T)}$ which does preserve it. In other words,
denoting any of the invariant objects $\eta$, $J$ or $\Ox$ by $\xi$ we
have
\begin{equation}
  \label{compatible}
  \nabla^{\rm (T)}\xi =0 \; .
\end{equation}
Any connection, and in particular $\nabla^{\rm (T)}$ defined above, can
be expressed in terms of the Levi-Civita connection $\nabla$ as
\begin{equation}
  \label{nablaT}
  \nabla^{\rm (T)}_m = \nabla_m + \k_m \ ,
\end{equation}
where $\k_m$ are matrices whose entries constitute the contorsion
tensor $\k_{mnp}$. Unlike the Levi-Civita connection, this connection
has a torsion $T_{mnp}=\k_{[mn]p}$.
Note that the contorsion tensor is anti-symmetric in its last two indices
and can be thought of as a one-form taking values in ${\rm so}(6)$, the
Lie-algebra of ${\rm SO}(6)$. Thus, we can decompose it under the
${\rm  SU}(3)$ structure group as
\begin{equation}
  \k_m = \k^0_m+\k^{{\rm su}(3)}_m\; ,
\end{equation}
where $\k^{{\rm su}(3)}_m$ takes values in ${\rm su}(3)={\bf 8}$, the
Lie-algebra of ${\rm SU}(3)$ and $\k^0_m$ takes values in the
complement ${\rm su}(3)^\perp={\bf 1}\oplus{\bf 3}\oplus\bar{\bf 3}$ of ${\rm
su}(3)$ within ${\rm so}(6)$. The action of $\k^{{\rm su}(3)}$ on the
${\rm SU}(3)$ invariant tensors $\x$ vanishes and, hence, the
left-hand side of the compatibility condition~\eqref{compatible} only
depends on $\k^0$ which is called the ``intrinsic contorsion''.
This intrinsic contorsion can be used to classify ${\rm SU}(3)$
structures and it is useful, in this context, to analyze its
${\rm SU}(3)$ representation content. From what has been said
above, the intrinsic contorsion $\k^0$ is an element of the ${\rm SU}(3)$
representation
\begin{equation}
  \label{Tclasses}
  ({\bf 3} \oplus \bar{\bf 3}) \otimes ({\bf 1} \oplus{\bf 3} \oplus
  \bar{\bf 3}) = ({\bf 1} \oplus {\bf 1}) \oplus ({\bf 8} \oplus{\bf
    8}) \oplus ({\bf 6} \oplus\bar{\bf 6}) \oplus ({\bf 3} \oplus
  \bar{\bf 3}) \oplus ({\bf 3}\oplus\bar{\bf 3})'\; .
\end{equation}
The five terms on the right-hand side of this relation correspond to
the five torsion classes~\cite{CS}, denoted by ${\cal W}_1,\dots
,{\cal W}_5$, of six-dimensional manifolds with ${\rm SU}(3)$
structure. These classes are a useful tool to characterize the
intrinsic torsion and the associated ${\rm SU}(3)$ structure. The
intrinsic torsion can also be read off from the exterior derivatives
$dJ$ and $d\O$ since Eq.~\eqref{compatible} implies that
\begin{eqnarray}
  (dJ)_{mnp} &=& 6 \; \k^0_{[mn}{}^r J_{r|p]} \label{dJ}\\
  (d\Ox )_{mnpq} &=& 12 \; \k^0_{[mn}{}^r \Ox_{r|pq]}\; .\label{dO}
\end{eqnarray}
Therefore, a practical way to specify the intrinsic torsion of an
${\rm SU}(3)$ structure is to explicitly write down expressions for
$dJ$ and $d\O$. As can be seen from Eqs.~\eqref{dJ} and \eqref{dO},
these expressions contain information about various of the five torsion
classes, namely
\begin{equation}
 dJ\in {\cal W}_1\oplus{\cal W}_3\oplus{\cal W}_4\; ,\qquad
 d\O\in{\cal W}_1\oplus{\cal W}_2\oplus{\cal W}_5\; . \label{JOtorsion}
\end{equation}

It will turn out that the first torsion class $\W_1$ plays a
special role in the case we address in this paper. Thus we define the
corresponding contorsion to be
\begin{equation}
  \label{k1}
  \k_{mnp}^{} \Big|_{\W_1} = \k_1 \Ox_{mnp} + \bar \k_1 \bar
  \Ox_{mnp} \; ,
\end{equation}
where $\k_1$ is given by
\begin{equation}
  \k_1 = \frac{i \int\sqrt{g}\, (dJ)_{mnp} \bar \Ox^{mnp}}{6
         \int\sqrt{g}\, \Ox_{mnp} \bar \Ox^{mnp}} \ ,
\end{equation}
and we have used Eq.~\eqref{dJ}.

\vspace{0.4cm}

In this paper, we are interested in a more special class of manifolds
with ${\rm SU}(3)$ structure, namely half-flat manifolds. They are
defined as six-dimensional ${\rm SU}(3)$ structure manifolds with
the invariant forms $J$ and $\O$ satisfying
\begin{equation}
 d\O_-=0\; ,\qquad dJ\wedge J=0\; ,\label{halfflat}
\end{equation}
where $\O_-$ is the imaginary part of $\O$. Comparison with
Eq.~\eqref{JOtorsion}  reveals~\cite{CS} that these conditions are
equivalent to vanishing torsion classes ${\cal W}_1^-$, ${\cal W}_2^-$
(these being the imaginary parts of the classes ${\cal W}_1$ and
${\cal W}_2$), ${\cal W}_4$ and ${\cal W}_5$.

\vspace{0.4cm}

The specific half-flat manifolds considered in this paper arise in the
context of mirror symmetry with NS-NS flux~\cite{GLMW}. Let us
briefly review how this comes about. Consider a mirror pair $X$ and
$Y$ of Calabi-Yau manifolds and introduce a standard symplectic basis
$(\tilde{\a}_I,\tilde{\b}^I)$, where $I=0,\dots h^{2,1}(X)$, of the
third cohomology on $X$. We start with, say, type IIB on $X$ in the
presence of NS-NS flux $\tilde{H}=e_i\tilde{\b}^i$, where
$i=1,\dots, h^{2,1}(X)$
and $\z=(e_i)$ are real flux parameters. Is there any compactification
on the IIA side which is mirror-symmetric to this configuration? Evidence for
this has been presented in Ref.~\cite{GLMW,GM} and it has been shown
that the mirror configuration is given by IIA on a half-flat manifold
$\hat{Y}_\z$, closely related to the original mirror Calabi-Yau $Y$,
but without NS-NS flux. Moreover, it has been argued that the moduli
spaces of metrics on $Y$ and $\hat{Y}_\z$ are identical for all values
of the flux $\z$.

Let us now describe the structure of these half-flat mirror manifolds
$\hat{Y}_\z$ in more detail. Matching of the moduli spaces of metrics,
together with the correspondence between metrics and ${\rm SU}(3)$
structures implies that the forms $J$ and $\O$ have expansions
\begin{eqnarray}
 J &=& v^i\o_i \label{J}\\
 \O &=& z^A \a_A -\cG_A\b^A \label{O}
\end{eqnarray}
similar to the ones on the associated Calabi-Yau manifold $Y$. Here
$(\o_i)$, where $i,j,\dots = 1,\dots ,h^{1,1}(Y)$ are $(1,1)$-forms
and $(\a^A,\b_A)$, where $A,B,\dots = 0,\dots ,h^{2,1}(Y)$, are
three-forms, suitable for the expansion of $J$ and $\O$ while the
coefficients $v^i$ and $z^A$ are the analog of K\"ahler and complex
structure moduli. For simplicity, we will continue to use Calabi-Yau
terminology and refer to K\"ahler and complex structure moduli and
moduli spaces, although the manifolds $\hat{Y}_\z$ are generally
neither K\"ahler nor complex. The three-forms $(\a^A,\b_A)$ satisfy
the standard normalizations
\begin{equation}
  \label{norm}
  \int_{\hat{Y}_\z} \a_A \wedge \b^B = \d^B_A \; ,\qquad
  \int_{\hat{Y}_\z} \a_A \wedge \a_B = \int_{\hat{Y}_\z} \b^A \wedge \b^B
  = 0 \; ,
\end{equation}
and we also introduce dual four-forms $\tilde{\o}^i$ such that
\begin{equation}
  \int_{\hat{Y}_\z} \o_i \wedge \tilde{\o}^j = \d_i^j \; .
\end{equation}
So far, all relations are identical to the corresponding Calabi-Yau
ones. However, unlike in the Calabi-Yau case the forms
$\o_i$ and $(\a_A,\b^B)$ are not all closed and, in particular, do not
form a basis of the second and third cohomology. Rather, as shown in
Ref.~\cite{GLMW}, mirror symmetry requires them to satisfy the
differential relations
\begin{equation}
 d\a_0 = e_i\tilde{\o}^i\; ,\qquad
 d\a_a = 0\; ,\qquad d\b^A = 0\; ,\qquad
 d\o_i = e_i\b^0\; ,\qquad
 d\tilde{\o}^i = 0\; , \label{drel}
\end{equation}
where we have introduced indices
$a,b,\dots = 1,\dots ,h^{2,1}(Y)$. The real parameters $e_i$ are
precisely the NS-NS flux parameters on the mirror side mentioned
earlier and they encode the degree to which the half-flat mirror
manifold $\hat{Y}_\z$ ``deviates'' from the associated Calabi-Yau
manifold $Y$. Using the above relations together with the
expansions~\eqref{J} and \eqref{O} for $J$ and $\O$ it is easy to show
that
\begin{eqnarray}
  \label{dJhf}
 dJ &=& v^ie_i\b^0 \\
 d\O &=& e_i\tilde{\o}^i\; .
\end{eqnarray}
As discussed, the right-hand sides of these relations specifies the
intrinsic torsion and the ${\rm SU}(3)$ structure of the manifolds
$\hat{Y}_\z$. Comparison with the conditions~\eqref{halfflat} shows that
they are indeed half-flat manifolds.

The point of view taken in this paper is that mirror symmetry with
NS-NS flux provides us with a practical ``definition'' of the
half-flat manifolds $\hat{Y}_\z$ as well as with a set of relations
which allows us to deal with them. The evidence for mirror
symmetry with NS-NS flux was obtained in the context of IIA and
IIB supergravity~\cite{GLMW} and one should, hence, expect the
above relations to be valid only in the large complex structure
limit. We will, therefore, work in this limit, in addition to the
large radius limit in K\"ahler moduli space which is mandatory
whenever supergravity theories are considered. In this paper we
are not interested primarily in mirror symmetry itself but in
using the so-defined manifolds in the context of the heterotic
string. We can see a number of advantages in this method compared
to, for example, working with the heterotic string on general
manifolds of ${\rm SU}(3)$ structure or even general half-flat
manifolds. Firstly, mirror symmetry strongly suggests that the
manifolds $\hat{Y}_\z$ actually exist although we are not aware
that examples of these manifolds have been explicitly constructed
except for the non-compact cases considered in Ref.~\cite{BDKT}.
Secondly, we have a relatively simple and explicit set of
differential relations, describing these half-flat mirror
manifolds, which facilitates concrete calculations. And finally,
from mirror symmetry one expects a half-flat mirror manifold
$\hat{Y}_\z$ for each Calabi-Yau space $X$ with a mirror $Y$ and
each set of flux parameters $\z$. This means we are dealing with a
large class of manifolds closely related to Calabi-Yau manifolds.
Hopefully this allows one to keep some of the phenomenologically
attractive features of heterotic Calabi-Yau
compactifications~\cite{CHSW}  while gaining additional benefits,
for example in terms of moduli stabilization through flux.


\section{Heterotic on half-flat: the bosonic action to lowest order in $\a '$}

We will now carry out the dimensional reduction of the heterotic
string on the half-flat mirror manifolds~\footnote{For convenience, we
will drop the index $\z$ on $\hat Y_\z$ from here on.} $\hat{Y}$
described in the previous section. We only consider the reduction of
the bosonic part of the action which should be sufficient to obtain
all the relevant information about the four-dimensional effective
action. However, the reduction of some of the fermionic terms provides
some additional insights and confirmation of the bosonic
results and we will come back to this in the following section. For
now, we restrict the calculation to lowest (zeroth) order in $\a '$
which, in particular, means we will not deal with gauge fields at this
stage. We will discuss the inclusion of gauge fields later.


\subsection{The reduction}

We would now like to compactify the zeroth order bosonic action~\eqref{SB}
on a half-flat mirror manifold $\hat{Y}$. As usual in flux
compactifications, the collective modes are taken to be the same as
for the corresponding case without flux, that is, as for the reduction
on the associated Calabi-Yau manifold $Y$, in our case. This approach
is in line with the earlier statement that the moduli spaces of the
half-flat mirror manifolds $\hat{Y}$ and the associated
Calabi-Yau
manifolds $Y$ are identical. Of course, one expects the flux to induce
a low-energy potential and, potentially, masses for some of the
previously massless fields. The idea will be that this ``flux'' scale
is sufficiently lower than the string and Kaluza-Klein scales. Only
then can heavy string/Kaluza-Klein modes be neglected while modes
acquiring masses from flux effects can be kept. This can be achieved
by sufficiently small flux parameters $e_i$ and/or large radii of the
internal manifold. At any rate, this separation of scales can be
consistently checked once the low-energy potential has been
computed. Although one expects the flux parameters $e_i$ to be
quantized (since the NS-NS flux of the mirror is quantized) we will here
work in a supergravity approximation and view them as continuous
parameters. We also adopt the general principle that our low-energy
effective theory should reduce to the standard one, obtained from
the reduction on the associated Calabi-Yau manifold $Y$, in the limit
of vanishing flux parameters, $e_i\rightarrow 0$.

\vspace{0.4cm}

We split 10-dimensional coordinates as $(x^M)=(x^\m ,x^m)$ with
external indices $\m ,\n ,\dots = 0,1,2,3$ and internal indices
$m,n,\dots = 4,\dots 9$. The 10-dimensional metric for our reduction
then takes the form
\begin{equation}
  \label{g10}
  ds_{10}^2 = e^{2\f} g_{\m\n}dx^\m dx^\n +g_{mn}dx^mdx^n\; ,
\end{equation}
where $g_{mn}$ is the metric on the half-flat mirror manifold $\hat{Y}$
induced by the ${\rm SU}(3)$ structure and $g_{\m\n}$ is the
four-dimensional metric. We have also introduced the zero mode
\begin{equation}
  \label{4ddil}
  \f = \hat{\f} - \frac12 \ln{\cal V}\; ,
\end{equation}
of the dilaton where ${\cal V}$ is the volume
\begin{equation}
  {\cal V}=\frac{1}{v}\int_{\hat{Y}}d^6x\sqrt{g}
\end{equation}
of the internal space $\hat{Y}$, measured relative to a fixed
reference volume $v$. The dilaton factor in front of the
four-dimensional part of the metric~\eqref{g10} has been chosen so
that we arrive at a canonically normalized Einstein-Hilbert term in
four dimensions. As we have already explained, the moduli space of
internal metrics $g_{mn}$
on $\hat{Y}$ is parameterized by K\"ahler moduli $v^i$, where
$i,j,\dots = 1,\dots ,h^{1,1}(Y)$ and complex structure moduli $z^a$,
where $a,b,\dots = 1,\dots ,h^{2,1}(Y)$. More specifically, we can
write the following standard equations for the deformations of the
metric
\begin{equation}
  \begin{aligned}
    \d g_{\a\bar{\b}} = & \ -i\o_{i\a\bar{\b}} \ \d v^i \\
    \d g_{\bar{\a} \bar{\b}} = & \ -\frac{1}{||\Ox||^2}
    \bar{\O}_{\bar{\a}}{}^{\g \d} (\chi_a)_{\gx \dx \bar{\bx}} \ \dx z^a\; ,
  \end{aligned}
\end{equation}
where we have introduced a set of $(2,1)$--forms $\c_a$ and
holomorphic (anti-holomorphic) indices $\a ,\b ,\dots$
($\bar{\a},\bar{\b},\dots$) on the internal space. Finally, we
have the following zero mode expansion for the NS-NS two-form
\begin{eqnarray}
  \hat{B} &=& B + b^i \o_i \label{Bansatz} \\
  \hat{H} &=& H + db^i \wg \o_i + (b^i e_i) \b^0\; , \label{Hansatz}
\end{eqnarray}
where $B$ is a four-dimensional two-form with field strength $H=dB$
and $b^i$ are $h^{1,1}(Y)$ real scalar fields.
Note that the last term in the Ansatz~\eqref{Hansatz} for the field
strength $\hat{H}$ is new compared to the Calabi-Yau case and results,
via Eq.~\eqref{drel}, from the fact that the $(1,1)$--forms $\o_i$ are no
longer closed. This term does have the form of (a particular type of)
$H$--flux, although it should be kept in mind that it originates
from the intrinsic ``flux'' encoded in the half-flat mirror manifolds.
For now we will not include genuine $H$--flux into the calculation
but defer this until later in the section.

\vspace{0.4cm}

Inserting the Ansatz~\eqref{g10}--\eqref{Hansatz} into the
10-dimensional bosonic action~\eqref{SB} one finds, after integrating
over the internal space
\begin{eqnarray}
  S_4 &=& -\frac{1}{2\k_4^2} \int \Big\{R\star 1 + 2 d \phi \wg \star
    d \phi +\frac12 d a \wedge \star d a + 2 \gK_{i \bar{j}}
    d t^i \wedge \star d\bar t^{\bar{j}}\nonumber\\
  &&\qquad\qquad + 2 \gcs_{a \bar b} d z^a \wedge \star
    d \bar z^{\bar b} + 2\k_4^2V *1\Big\} \; ,\label{S4}
\end{eqnarray}
with the four-dimensional Newton constant $\k_4^2=\k_{10}^2/v$ and the
scalar potential
\begin{equation}
  V = 4\k_4^{-2} e^{2\phi + \KK + \Kcs}
  \left[e_i e_j \gKi^{ij} + 4(e_i b^i)^2 \right]\; .\label{V}
\end{equation}
The complex K\"ahler moduli $t^i$ are defined by
\begin{equation}
  \label{tdef}
  t^i = b^i+iv^i\; ,
\end{equation}
and the four-dimensional two-form $B$ has been dualized to the scalar
$a$. The K\"ahler and complex structure moduli space metrics are
defined as usual by
\begin{eqnarray}
  \gK_{ij} &=& \frac{1}{4v {\cal V}} \int_{\hat{Y}} \o_i \wg
  \star \o_j \\
  \gcs_{a\bar{b}} &=& -\frac{\int_{\hat{Y}} \c_a \wedge
  \bar{\c}_{\bar{b}}}{\int_{\hat{Y}} \O \wedge \bar{\O}} \; ,
\end{eqnarray}
with inverse metrics $\gKi^{ij}$ and $\gcsi^{a\bar{b}}$ and
associated K\"ahler potentials
\begin{equation}
  \label{KKcs}
  \begin{aligned}
    \KK = & -\ln (8{\cal V}) \\
    \Kcs = & -\ln\left(
    i\int_{\hat{Y}}\O\wedge\bar{\O}\right)\; .
  \end{aligned}
\end{equation}
In this calculation, we have used the following result for the
integrated scalar curvature of half-flat mirror manifolds
\begin{equation}
  \label{VR}
  \int_{\hat{Y}}\sqrt{g}\, R_{\rm hf} = v\exp\left(\Kcs\right)
  e_ie_j \gKi^{ij}\; ,
\end{equation}
which was proven in Ref.~\cite{GLMW}, as well as the special geometry
relations~\eqref{ABC} and \eqref{M-1} in order to evaluate the
integral $\int_{\hat{Y}}\b^0\wedge\star\b^0$. The two contributions to
the four-dimensional potential~\eqref{V} originate from this
non-vanishing scalar curvature and the additional term in the Ansatz
for the NS-NS 3-form field strength $\hat{H}$ in Eq.~\eqref{Hansatz},
respectively.


\subsection{Four-dimensional supergravity}
\label{relsugra}

The four-dimensional action derived in the previous subsection should be
the bosonic part of an $N=1$ supergravity theory. We would now like to make
this explicit comparing it to the standard $N=1$ supergravity
action~\cite{WB}.

The kinetic terms in~\eqref{S4} are easy to deal with since they are identical
to the ones arising in standard Calabi-Yau compactifications. We introduce
chiral superfields $S$, $T^i$ and $Z^a$ satisfying
\begin{eqnarray}
  S\left|\right. &=& a+ie^{-2 \f} \label{Sdef}\\
  T^i\left|\right. &=& t^i \label{Tdef}\\
  Z^a\left|\right. &=& z^a\; , \label{Zdef}
\end{eqnarray}
where the bar denotes the lowest component of the multiplet. Then the
K\"ahler potential reproducing the kinetic terms in Eq.~\eqref{S4} can be
written as
\begin{equation}
 K = \k_4^{-2}\left(K^{({\rm S})}+\KK+\Kcs\right)\; ,\label{K}
\end{equation}
where
\begin{equation}
 K^{({\rm S})} = -\ln\left( i(\bar{S}-S)\right) \; ,\label{KS}
\end{equation}
and $\KK$ and $\Kcs$ are given in \eqref{KKcs}. In order to
perform a concrete calculation one needs to express these
K\"ahler potentials in terms of the low energy fields. This is done
via holomorphic pre-potentials ${\cal F}$ and ${\cal G}$ and
the respective equations are given in \eqref{KKs} and \eqref{Kcs}.

\vspace{0.4cm}

Having fixed the K\"ahler potential and the superfields in terms of
component fields via Eqs.~\eqref{Sdef}--\eqref{Zdef} we now have to
check whether the potential~\eqref{V}, obtained from dimensional
reduction, can be reproduced from the standard supergravity expression
\begin{equation}
 V = \k_4^{-4}e^{\k_4^2K}\left(K^{X\bar{Y}}D_XWD_{\bar{Y}}\bar{W}-3\k_4^2|W|^2\right)\; ,
 \label{VSUGRA}
\end{equation}
for a suitable choice of superpotential $W$. In this expression, we have
used indices $X,Y,\dots$ to label all chiral superfields $(\F^X)=(S,T^i,Z^a)$
and $D_X$ denotes the K\"ahler-covariant derivative defined by
\begin{equation}
  D_X W = \partial_X W + \k_4^2K_X W\; .
\end{equation}
Further $K^{X\bar{Y}}$ is the inverse of the K\"ahler metric $K_{X\bar{Y}}$.

The potential~\eqref{V} is quadratic in the axionic fields $b^i$ which
are part of the chiral multiplets $T^i$. This suggest that the superpotential
may be a linear function in the fields $T^i$. In fact, we claim that $W$ is
given by
\begin{equation}
  W= \sqrt8\, e_i T^i \; . \label{W}
\end{equation}
Let us now verify this claim. We first note that, using the
expression~\eqref{K} for the K\"ahler potential, the pre-factor in the
reduction potential~\eqref{V}
can be re-written as
\begin{equation}
  4\exp\left(2\phi+\KK+\Kcs\right)=8e^{\k_4^2K}\; .
  \label{prefactor}
\end{equation}
This correctly matches the $e^{\k_4^2K}$ pre-factor of the
supergravity potential~\eqref{VSUGRA}. With the
superpotential~\eqref{W}, the various K\"ahler-covariant derivatives
are given by
\begin{eqnarray}
 D_S W &=& -\frac{1}{2}e^{2\f} W \\
 D_i W &=& \sqrt{8}e_i+\KK_i W \\
 D_a W &=& \Kcs_a W \; .
\end{eqnarray}
For the non-vanishing components of the K\"ahler metric we have
$K_{S\bar{S}}=e^{2\f}/4$, $K_{i\bar{j}}=\gK_{i\bar{j}}$
and $K_{a\bar{b}} = \gcs_{a\bar{b}}$. Using these, and
Eq.~\eqref{KKs} we find
\begin{eqnarray}
  D_S W D_{\bar{S}} \bar{W} K^{S\bar{S}} &=& |W|^2 \label{Spart}\\
  D_i W D_{\bar{j}} \bar{W} K^{i\bar{j}} &=& 8e_ie_j \gKi^{ij} -
  32(e_i v^i)^2 + 3|W|^2
 \label{Tpart}\\
 D_a W D_{\bar{b}}\bar{W}K^{a\bar{b}} &=& 3|W|^2\; . \label{Zpart}
\end{eqnarray}
In the second line, we have used \eqref{uf} which holds for special
geometries with a cubic pre-potential.
The result in the third line can be
proved using a similar cubic pre-potential~\eqref{csprep} for the complex
structure moduli which is justified in the large complex structure limit.
Inserting the relations~\eqref{prefactor} and~\eqref{Spart}--\eqref{Zpart}
into the supergravity potential~\eqref{VSUGRA}, using the explicit
form~\eqref{W} of $W$ we indeed correctly reproduce the
potential~\eqref{V} obtained from the reduction.

\vspace{0.4cm}

To summarize our results so far, we have derived, to lowest order in
$\a '$, the bosonic part of the four-dimensional effective action of
the heterotic string on half-flat mirror manifolds $\hat{Y}$. We have
shown that this action is indeed the bosonic part of a
four-dimensional $N=1$ supergravity theory with K\"ahler
potential~\eqref{K} and superpotential~\eqref{W}. This latter
statement has been proved for large complex structure since we have
used the relation~\eqref{Zpart} which, as far as we know, only holds
in this limit. Given that the relations which define the half-flat
mirror manifolds can only be expected to hold for large complex
structure this is perhaps not surprising. However, our result
indicates that the definition of the half-flat mirror manifolds indeed
has to be modified away from the large complex structure limit.


\subsection{Including H-flux}

Our previous calculation can be generalized by adding an arbitrary
three-form $H_{\rm flux}$, harmonic on the internal space, to the
Ansatz~\eqref{Hansatz} for the NS-NS field strength $\hat{H}$. In the
analogous Calabi-Yau case, the forms $(\a_A,\b^B)$ constitute a basis
of harmonic three-forms and the most general NS-NS flux is simply
given by an arbitrary linear combination of these forms. Here, we have
to be more careful. From Eq.~\eqref{drel} we know that $\a_0$ is not
even closed which means it does not define a cohomology class. All
other forms $(\a_a ,\b^b)$ are closed but not necessarily
co-closed. However, we know that
\begin{equation}
   \label{ds}
  d\star\a_a = {\cal O}(e_i)\; ,\qquad d\star\b^b = {\cal O}(e_i)\; ,
\end{equation}
since these forms are harmonic in the Calabi-Yau limit $e_i\rightarrow 0$.
Hence, the forms $(\a_a,\b^b)$ define cohomology classes and they differ
from the harmonic representative by exact forms of the order $e_i$.
This understood, we write the following Ansatz for the NS-NS flux
\begin{equation}
  H_{\rm flux} = \e_A \bx^A + \m^A \ax_A\; ,\label{Hflux}
\end{equation}
where
\begin{equation}
  \label{NSflux}
  \e_A = (0, \, \e_a) \; , \qquad \mu^A= (0, \, \mu^a) \; .
\end{equation}
We have allowed indices in \eqref{Hflux} to run over all values to
keep expressions covariant but we have set $\m^0=0$ in accordance with
the above discussion. Also we note that dealing with the flux
parameter $\e_0$ is a bit more subtle as it was argued in \cite{GLMW}
that it reproduces the mirror of the zero-NS-flux. For this reason, we
have also set $\e_0 = 0$. However, all other flux parameters $(\e_a,
\mu^a)$ are kept arbitrary. The so-defined NS-NS flux satisfies
\begin{equation}
  dH_{\rm flux} = 0\; ,\qquad d\star H_{\rm flux} = (\mbox{second order
    in flux})\; .
\end{equation}
For the second relation, we have used Eq.~\eqref{ds} and, here and
in the following, ``$n^{\rm th}$ order in flux'' refers to a
quantity proportional to a product of $n$ of the flux parameters
$e_i$, $\e_a$ or $\m^a$.

\vspace{0.4cm}

We would now like to repeat our reduction of the lowest order
bosonic action~\eqref{SB}, using the
Ansatz~\eqref{g10}--\eqref{Hansatz}, but modifying the expression
for $\hat{H}$ by adding to it the NS-NS flux~\eqref{Hflux}. The
kinetic terms are, of course, unmodified by the additional
NS-NS-flux and the four-dimensional effective action is still of
the form~\eqref{S4}, where only the potential $V$ has a different
form. Combining our earlier expression~\eqref{Hansatz} for the field
strength $\hat{H}$ with the $H$-flux~\eqref{Hflux} we have
\begin{equation}
  \hat{H} = H + db^i \wg \o_i + (b^i e_i) \b^0 + \e_a \bx^a + \m^a
  \ax_a\ .
\end{equation}
The contribution to the potential which originates from the
non-vanishing scalar curvature~\eqref{VR} of the half-flat mirror
manifolds remains the same. However, we have to consider the
additional terms which arise from this new form of $\hat{H}$
when inserted into the form field kinetic term. To do this,
we note, the term proportional to $\bx^0$ in the above expression
looks like an ordinary $H$--flux and can be treated on the
same footing. To this end, we define the modified flux parameters
$\tilde \e_A=(e_i b^i , \; \e_a)$. With these, the potential takes the
form
\begin{equation}
  \label{Vflux}
  V = 4 e^{2 \phi + \KK + \Kcs} e_i e_j \gKi^{ij} -
  2e^{2\phi + \KK} (\tilde \e_A + \mu^c \cM_{cA})
  (\mathrm{Im} \cM)^{-1 \; AB} (\tilde \e_B + \mu^d \bar \cM_{dB}) \; .
\end{equation}
To obtain the last term we have used \eqref{ABC} and \eqref{A-N} and
the matrix ${\cal M}$ is defined in Eq.~\eqref{eq:M}. Since we have
neglected second order flux terms in $H_{\rm flux}$ this potential is
correct up to quadratic terms in the flux and there are possible
corrections of cubic and higher order in flux which we have not
calculated. Let us also note that despite the explicit minus sign
which appears in the above formula, this potential is manifestly
positive definite as the matrix $(\mathrm{Im}(\cM))^{-1}$ is negative
definite. When deriving the potential from the superpotential, this
feature will arise from the no-scale structure which
annihilates the negative contribution in \eqref{VSUGRA}. We note that
Eq.~\eqref{Vflux} reduces to the previous formula~\eqref{V} for the
potential in the absence of $H$--flux by setting $\e_a=0$ and
$\m^a=0$, remembering that $\tilde \e_0 = e_i b^i$.

\vspace{0.4cm}

As before, it has to be checked that the above result can be embedded into
four-dimensional $N=1$ supergravity. Since the kinetic terms are
unmodified, the definition of superfields is still given by~\eqref{Sdef}--\eqref{Zdef}
and the K\"ahler potential is the standard one, Eq.~\eqref{K}. Given these results,
is the modified potential~\eqref{Vflux} of the supergravity form~\eqref{VSUGRA}
for a suitable superpotential $W$? It is shown in Appendix~\ref{appB} that this
is indeed the case, provided one is working in the large complex structure limit.
The superpotential then reads
\begin{equation}
 W= \sqrt 8 (e_iT^i + \e_a Z^a + \m^a \cG_a )\; , \label{W1}
\end{equation}
with arbitrary flux parameters $e_i$, $\e_a$ and $\m^a$.


\section{Gravitino mass and the superpotential}

In this section we propose another approach to compute the
superpotential which will turn out to be more suitable for further
generalizations and for obtaining some more insight when $\alpha'$
corrections are taken into account. Previously, we have derived the
moduli superpotential by dimensional reduction of the bosonic action
and by comparing the result with the standard form of four-dimensional
$N=1$ supergravity.  However, there is also a more direct method using
Gukov's formula~\cite{GVW,SG} which, in the appropriate form, has led
to the correct result for a number of different
compactifications.  In this section, we are going to
explore this second approach and its relation to the results of the
previous section, for the case of heterotic string on half-flat mirror
manifolds.

We will proceed in two steps. Firstly, we will derive the appropriate
version of Gukov's formula from the four-dimensional gravitino mass
term which we obtain as a dimensional reduction of the appropriate
terms in the 10-dimensional action, an approach also considered in
Refs.~\cite{BBHL,BBDG}. As we will see, the resulting Gukov-type
formula applies to the heterotic string on all manifolds of ${\rm
SU}(3)$ structure and is valid to first order in $\a '$. As a second
step, we then apply this general formula to our particular half-flat
mirror manifolds and show that it specializes to the
superpotential~\eqref{W1}, derived in the previous section.


\subsection{A Gukov formula from the gravitino mass term}

The mass term for the gravitino $\Psi$ in four dimensions is given by
\begin{equation}
  \label{massterm}
  S_{\Psi ,{\rm mass}} = -\frac12 \int_{M_4}d^4x \sqrt{-g} \left\{ M_{3/2}
    \Psi_\m^\dagger \gx^0 \g^{\mu\nu}\Psi^*_\nu +\mbox{h.c.} \right\}\; ,
\end{equation}
where the four-dimensional gamma matrices $\g_\m$ are chosen
to be real and the chirality matrix
\begin{equation}
 \g = -\frac{i}{4!}\e^{\m\n\r\s}\g_{\m\n\r\s}\; ,
\end{equation}
is purely imaginary. In the context of $N=1$ four-dimensional supergravity
the gravitino mass can be written as
\begin{equation}
  \label{M32}
  M_{3/2} = \exp (\k_4^2K/2)W\; ,
\end{equation}
and the invariant function $G=K+\ln |W|^2$ can be computed from the
gravitino mass using the relation
\begin{equation}
 \label{G}
 e^G=|M_{3/2}|^2\; .
\end{equation}
If the K\"ahler potential has been computed independently or the holomorphic
part of $M_{3/2}$ can be identified then the superpotential can be
obtained directly from $M_{3/2}$. We are now going to apply these facts to
the gravitino mass term which descends from the 10-dimensional theory.

\vspace{0.4cm}

A quick inspection of the ten dimensional action~\eqref{SF} reveals
which parts potentially contribute to the gravitino mass terms in four
dimensions. The most obvious one is the ``flux'' term $\Psi_M
\G^{MNPQR} \Psi_N H_{PQR}$. This term was also considered in
Ref.~\cite{BBHL,BBDG,HPN} and, as we will show, it gives
rise to the 
well known superpotential $W \sim \int H \wedge\O$ which was proposed
in Refs.~\cite{GVW,SG}. This result for $W$ is definitely correct for
Calabi-Yau manifolds, but if the internal manifold has only $SU(3)$
structure there will be a further contribution from the gravitino
kinetic term in ten-dimensions. This additional contribution will turn
out to be proportional to the first torsion class, ${\cal W}_1$, of
the ${\rm SU}(3)$ structure manifold.  The reason this term appears in
four dimensions is that on such manifolds the globally defined spinor
$\eta$ is no longer covariantly constant with respect to the
Levi-Civita connection.

\vspace{0.4cm}

Let us now see how this works in detail. We first have to decompose the
10-dimensional gamma matrices
\begin{equation}
  (\g_M) = (\g_\m \otimes{\bf 1}, \g \otimes \g_m) \; .
  \label{gxdec}
\end{equation}
Note that we have chosen the four-dimensional gamma matrices, $\g_\m$,
real and the six-dimensional ones, $\g_m$, imaginary so that the above
decomposition leads to real 10-dimensional gamma
matrices. Furthermore, we have to decompose the 10-dimensional
gravitino $\hat{\Psi}_M$ in a way compatible with its Majorana-Weyl
nature. The unique possibility, up to overall rescalings, for the case
of a manifold with $SU(3)$ structure is
\begin{equation}
  \label{grdec}
  \hat \Psi_M = e^{\frac\phi2} \Big(\psi_M \otimes \eta + \psi_M^*
  \otimes \eta^* \Big) \, .
\end{equation}
where $\psi_M$ is a four dimensional Weyl spinor of positive
chirality. We recall that $\eta$ is the six-dimensional globally
defined Weyl spinor which exists on manifolds with ${\rm SU}(3)$
structure.  The external components $\psi_\m$ correspond to the
four-dimensional gravitino while $\psi_m$ represent spin $1/2$
fields. In fact, in order not to have cross kinetic terms between the
gravitino and the spin $1/2$ fields one needs to redefine $\psi_\mu$
by some particular combination of $\psi_m$. However, this subtlety
does not effect the gravitino mass which can be read off as the
coefficient of the term $\frac12 \psi^\dagger_\mu \gx^0 \gamma^{\mu
\nu}\psi^*_\nu$.  On the other hand the normalization of the gravitino
field is important since its kinetic term needs to be in canonical
form in order to read off the correct gravitino mass. For this reason
we have chosen the overall factor $e^{\phi/2}$ in Ansatz~\eqref{grdec} and
one can easily check that this leads to the correct kinetic term for
the gravitino in four dimensions. Let us quickly sketch how this
works. Inserting \eqref{grdec} and \eqref{gxdec} into the
ten-dimensional kinetic term from \eqref{SF} and keeping only the
terms involving the four-dimensional space-time indices we obtain
\begin{equation}
  \overline{\hat \Psi}_\m \Gx^{\mu \nu \rho} D_\nu \hat \Psi_\rho = e^\phi\left(
  \overline \psi_\mu \gx^{\mu \nu \rho} D_\nu \psi_\rho \;
  (\eta^\dagger \eta) + \psi_\mu^T \gx^{\mu \nu \rho} D_\nu \psi^*_\rho
  \; (\eta^T \eta^*)\right) \; .
\end{equation}
Note that due to our conventions the above terms are the only
combinations which survive as $\eta^T \eta = \eta^\dagger \eta^*
\equiv 0$. Also recall that we have normalized the spinor $\eta$
requiring $\eta^\dagger \eta =1$ so that the terms above do not
depend on the internal manifold. Consequently the integration over the
six-dimensional space will only produce a volume factor which combines
with the dilaton factor in Eq.~\eqref{SF} into the four dimensional
dilaton \eqref{4ddil}. Finally, taking into account the rescaling of
the space-time metric \eqref{g10} we obtain for the four-dimensional
gravitino kinetic term
\begin{equation}
  \label{kin3/2}
  S_{\frac32 \mathrm kin} = \frac12 \int_{M_4}d^4x \sqrt {-g}
  \left[\overline \psi_\mu \gx^{\mu \nu \rho} D_\nu \psi_\rho +
    \psi^T_\mu \gx^{\mu \nu \rho} D_\nu \psi^*_\rho \right] \; ,
\end{equation}
which is indeed the correct kinetic term for the gravitino in four
dimensions \cite{WB}.

\vspace{.4cm}

Having normalized the gravitino field correctly we
can go ahead and derive the gravitino mass term. This can be done by
inserting the decompositions \eqref{gxdec} and \eqref{grdec}, along with
the Ansatz~\eqref{g10} for the 10-dimensional metric into the fermionic
action~\eqref{SF} and keeping the terms with two four-dimensional gamma
matrices and no space-time derivatives. Let us consider the two
relevant terms in \eqref{SF} separately, starting with the kinetic
term. We obtain
\begin{eqnarray}
  \label{Mkin}
  \overline{\hat \Psi}_\mu \G^{\mu n \nu} D_n \hat \Psi_\nu  & = &
  -(\psi_\mu \otimes \eta + \psi_\mu^* \otimes \eta^*)^T [\g^0
  \g^{\mu \nu} \otimes \g^n D_n] (\psi_\nu \otimes \eta + \psi_\nu^*
  \otimes \eta^*) \, .
\end{eqnarray}
From compatibility condition~\eqref{compatible} we know that the spinor
$\eta$ is covariantly constant with respect to the connection with torsion.
This implies
\begin{equation}
  D_n \eta - \frac14 \kappa_{npq} \g^{pq} \eta =0 \, ,
\end{equation}
which, applied to Eq.~\eqref{Mkin}, yields
\begin{equation}
  \overline{\hat \Psi}_\mu \G^{\mu n \nu} D_n \hat \Psi_\nu = \frac14
  \psi_\mu^\dagger \gx^0 \gx^{\mu \nu} \psi^*_\nu  \; (\eta^\dagger
  \gx^n \gx^{pq} \eta^*) \k_{npq} - \frac14 \psi_\mu^T \gx^0
  \gx^{\mu \nu} \psi_\nu  \; (\eta^T \gx^n \gx^{pq} \eta) \k_{npq} \; .
\end{equation}
As before, we have discarded terms like $\eta^\dagger \gx^n \gx^{pq}
\eta$ which vanish identically.  Moreover, the properties of
six-dimensional spinors and gamma matrices assure that only the
totally antisymmetric part of $\eta^\dagger \gx^n \gx^{pq} \eta^*$
survives. Using \eqref{Oeta} and taking care to include all the dilaton
factors in Eqs.~\eqref{g10} and \eqref{grdec}, we conclude that the
torsion contribution to the gravitino mass term can be written as
\begin{equation}
  \label{MT}
  M_{\frac{3}{2}}^{(T)} = \frac{e^\phi}{4 \mathcal V} \int_{\hat{Y}}
  \sqrt{g} \; \eta^\dagger \g^{npq} \eta^* \kappa_{npq} =
  \frac{i e^\phi \sqrt{8}}{4 {\mathcal V} ||\O ||}
  \int_{\hat Y} \sqrt{g} \; \O^{\ab \bb \cb} \kappa_{\ab \bb \cb} \; .
\end{equation}
With Eq.~\eqref{dJ} and the relation $J_m{}^n \Ox_{npq} = i \Ox_{mpq}$
one can also write the above expression in the following form
\begin{equation}
  \label{MTdJ}
   M_{\frac{3}{2}}^{(T)} =
  \frac{e^\phi \sqrt{8}}{24 {\mathcal V} ||\O ||}
  \int_{\hat Y} \sqrt{g} \; \O^{\ab \bb \cb} (dJ)_{\ab \bb \cb}
  = \frac{i e^\phi \sqrt{8}}{4 {\mathcal V} ||\O ||} \int_{\hat Y} \O
  \wg dJ \; .
\end{equation}
We recall from
Eq.~\eqref{Tclasses} that the torsion $\k$ decomposes into five
classes according to the various ${\rm SU}(3)$ representations it
contains. Evidently, contracting with $\O$ in the above relation
projects out the ${\rm SU}(3)$ singlet part which corresponds to
the torsion class ${\cal W}_1$.

For the $\hat{H}$-dependent term in the fermionic action~\eqref{SF}
the calculation is similar and was also discussed in
Refs.~\cite{BBHL,BCtin,BBDG,HPN}. One finds
\begin{equation}
  \label{gmH}
  \bar \Psi_\mu \G^{\mu npq \nu} \Psi_\nu \hat H_{npq} = -(\psi_\mu
  \otimes \eta + \psi_\mu^* \otimes \eta^*)^T [\g^0 \g^{\mu \nu} \g^5
  \otimes \g^{npq}] (\psi_\nu \otimes \eta + \psi_\nu^* \otimes
  \eta^*) \, ,
\end{equation}
and comparison with Eq.~\eqref{massterm} leads to the gravitino mass contribution
\begin{equation}
  \label{MH}
  M_{\frac{3}{2}}^{(H)} = -\frac{i e^{\phi} \sqrt8}{24 {\mathcal V}
    ||\O ||} \int_{\hat{Y}} \sqrt{g} \;\Omega^{\ab \bb \cb} \hat
  H_{\ab \bb \cb}
  = \frac{e^{\phi} \sqrt8}{4 {\mathcal V} ||\O ||} \int_{\hat Y} \O \wg H \; .
\end{equation}
Adding up the two contributions~\eqref{MTdJ} and \eqref{MH} one finds
for the gravitino mass
\begin{equation}
 M_{3/2} = \frac{\sqrt{8}e^\f}{4{\cal V}||\O ||}\int_{\hat{Y}}
       \O\wedge (H +idJ)\; .
 \label{Mtotal}
\end{equation}
From Eq.~\eqref{G} this determines the supergravity function $G$.

Of course we do not know the K\"ahler potential for general manifolds
with ${\rm SU}(3)$ structure. However, it is suggestive to identify
the integral in Eq.~\eqref{Mtotal} as the holomorphic part and, hence,
the superpotential and the pre-factor as the K\"ahler potential.
Accepting this we find by comparison with Eq.~\eqref{M32}
that~\footnote{A similar formula for the superpotential was first
proposed in Ref.~\cite{GLMW} in the context of type II theories. In
heterotic string compactifications it appears in Ref.~\cite{CCDL}, but
to our knowledge it was never derived before in a systematic way as we
do in this paper.}
\begin{equation}
  \label{Wmass}
  W \sim \int_{\hat{Y}}\O\wedge \left(\hat{H} + i dJ \right)\; .
\end{equation}
Note that $\hat{H} + idJ = d(\hat{B} + iJ)$ is precisely the holomorphic
combination which determines the scalar components~\eqref{Tdef} of the
superfields $T^i$. Equivalently, following the notation of Ref.~\cite{GMPT}
we can write the superpotential as
\begin{equation}
  W \sim H_1 + \k_1 \; ,
\end{equation}
where the ${\rm SU}(3)$ singlet component of the torsion, $\k_1$, was defined in \eqref{k1}.
Likewise, $H_1$ is the ${\rm SU}(3)$ singlet component of $\hat{H}$ defined by
\begin{equation}
  \hat{H}_{mnp}\left.\right|_{\rm singlet} = -6H_1(\Ox +\bar{\O})_{mnp}\; .
\end{equation}

Comparing again with \eqref{M32} we can argue that the pre-factor in
Eq.~\eqref{Mtotal} should determine the K\"ahler potential. Thus we
can write
\begin{equation}
  \label{Kmass}
  e^K \sim \frac{e^{2\phi}}{{\mathcal V}^2 ||\Ox||^2}
\end{equation}
We stress that one expects these results for $G$, $W$ and $K$ to be
valid for heterotic compactifications on all manifolds with ${\rm
SU}(3)$ structure. In addition, they hold up to
and including correction of order $\a '$ since the relevant
10-dimensional gravitino terms in Eq.~\eqref{SF} do not receive
corrections at this order. This latter fact can be illustrated for
standard Calabi-Yau compactifications. In this case it is
straightforward to show that the formula~\eqref{Wmass} correctly
reproduces the cubic gauge matter superpotential~\cite{GSW} which
arises at order $\a '$. We expect the relation~\eqref{Wmass} will be
quite useful when computing the gauge matter superpotential in more
general cases, such as for half-flat mirror manifolds.

\vspace{0.4cm}

\subsection{Application to half-flat mirror manifolds}

If the K\"ahler potential has been fixed by other means, the superpotential
can be obtained from Eq.~\eqref{Mtotal} exactly, including
the pre-factor. For example, using the Calabi-Yau K\"ahler potential~\eqref{K}
which, as we have seen, also applies to half-flat mirror manifolds one finds
\begin{equation}
  \label{Wmass1}
  W = \sqrt 8 \int_{\hat{Y}} \O\wedge\left(\hat{H} + i d J \right)\; ,
\end{equation}
It is now just a simple exercise to obtain the expression of the
superpotential in terms of the component fields in four dimensions.
We recall from Eqs.~\eqref{Hansatz} and \eqref{Hflux} that the
complete Ansatz for the NS-NS field strength, including NS-NS flux, is
given by
\begin{equation}
 \hat{H} = H + db^i\o_i + (b^ie_i) \bx^0 + e_A \bx^A + m^a \ax_a\; .
\end{equation}
Using the expansion of the $(3,0)$ form $\Ox$ in terms of the
complex structure moduli \eqref{O}, the particular expression for
$dJ$, \eqref{dJhf}, and the integration rules \eqref{norm} one
immediately obtains
\begin{equation}
  \label{W2}
  W= \sqrt 8 \left(e_iT^i + \e_a Z^a + \m^a {\cal G}_a\right) \; ,
\end{equation}
which precisely coincides with \eqref{W1}.
In summary, we have verified this result in two, largely independent
ways, namely by a reduction of the bosonic term and from a generalized
Gukov-type formula which we have derived from the gravitino mass term.


\section{Including gauge fields}

Let us now discuss some properties of the heterotic $E_8\times E_8$ string on
half-flat mirror manifolds at first order in $\a '$. At this order,
the Bianchi identity~\eqref{BI} for $\hat{H}$ receives its gauge field and
gravitational Chern-Simons correction and finding its solution becomes
a non-trivial task. With
\begin{equation}
 d\o_{\rm YM} = {\rm Tr}\left(\hat{F}\wedge\hat{F}\right)\; ,\qquad
 d\o_{\rm L} = {\rm tr}\left(\tilde{R}\wedge\tilde{R}\right)
\end{equation}
the Bianchi-identity leads to the well-known relation
\begin{equation}
  d\hat{H} = \frac{\a '}{4}\left({\rm tr} (\tilde{R} \wedge \tilde{R})
    - {\rm Tr}(\hat{F}\wedge \hat{F})\right)\; . \label{dH}
\end{equation}
It implies, as a condition for the Bianchi identity to be soluble,
that the right-hand side has to be cohomologically trivial and,
hence, that
\begin{equation}
 \left[{\rm tr}(\tilde{R}\wedge\tilde{R})\right] =
 \left[{\rm Tr}(\hat{F}\wedge\hat{F})\right]\; , \label{coh}
\end{equation}
where the bracket $[\dots ]$ denotes the cohomology class. Traditionally,
the way to satisfy this condition has been the standard embedding~\cite{CHSW}
although more general possibilities have been discussed in the
literature~\cite{DG,FMW,DLOW}.

\vspace{0.4cm}

Here, we will consider the simplest possibility, a generalization of
the standard embedding to our compactifications. Let us first recall
the standard Calabi-Yau case. The spin connection $\o^{({\rm CY})}_m$ of the
Calabi-Yau manifold $Y$ takes values in ${\rm SU}(3)$ which means its
non-vanishing components are of the form $\o_m^{({\rm
CY})\a\bar{\b}}$. The standard embedding then amounts to setting the
internal Yang-Mills connection equal to the Calabi-Yau spin
connection, that is
\begin{equation}
 \left. {A_m}^{\a\bar{\b}}\right|_{\mathrm{background}} = \o_m^{({\rm
 CY})\a\bar{\b}}\; .
\end{equation}
Here, the indices $(\a ,\bar{\b})$ on $A$ refer to an ${\rm SU}(3)$
sub-group of one of the $E_8$ factors of the gauge group. The trace of
the square of such an ${\rm SU}(3)$ generator in the adjoint of $E_8$
is $30$ times the trace of the square of an ${\rm SU}(3)$ generator in
the fundamental. With the definition of ${\rm Tr}$ as $1/30$ of the
trace in the adjoint of $E_8\times E_8$, this means that the standard
embedding indeed solves the cohomology constrain~\eqref{coh} and, even
more strongly, leads to the right-hand side of Eq.~\eqref{dH}
to vanish identically. Note that, at the level of background fields,
the internal part of $\hat{H}$ is vanishing so that the modification~\eqref{ot}
of the spin connection does not contribute for Calabi-Yau manifolds.
The surviving low-energy gauge group is the maximal commutant of
${\rm SU}(3)$ within $E_8\times E_8$ which is $E_6\times E_8$. In
addition, one obtains $h^{1,1}(Y)$ chiral multiplets in the
$\overline{\bf 27}$ of $E_6$ and $h^{2,1}(Y)$ chiral multiplets in the
${\bf 27}$ of $E_6$.

\vspace{0.4cm}

Can this picture be adapted to half-flat mirror manifolds? There
are two essential modifications. First of all, the spin connection
$\o^{({\rm hf})}_m$ of the half-flat manifold generally takes
values in ${\rm SO}(6)$ rather than ${\rm SU}(3)$. Secondly, the
internal background value of $\hat{H}$ is no longer vanishing due
to the additional term in Eq.~\eqref{Hansatz} and, if present,
$H$--flux in Eq.~\eqref{Hflux}. Therefore, we have to work
with the modified connection $\tilde{w}$ which is the correct
object that enters the Bianchi identity. From Eqs.~\eqref{ot},
\eqref{Hansatz} and \eqref{Hflux} it is given by
\begin{equation}
 \tilde{\o}_m{}^{np} = \o_m^{({\rm hf})np}+{\left( b^ie_i\b^0
                         +\e_a\b^a +\m^a\a_a\right)_m}^{np}\; .
\end{equation}
This connection still generically takes values in ${\rm SO}(6)$. The
generalization of the standard embedding to half-flat mirror manifolds
is then characterized by
\begin{equation}
 {A_m}^{np} = \tilde{\o}_m{}^{np}\; ,
\end{equation}
where the index pair $(np)$ on $A$ refers to an ${\rm SO}(6)$
sub-group of one of the $E_8$ gauge factors. The trace of the square
of an ${\rm SO}(6)$ generator in the adjoint of $E_8$ is still $30$
times that of the trace in the fundamental of ${\rm SO}(6)$ and,
hence, the above choice indeed provides a solution to the cohomology
constraint~\eqref{coh}. As for the Calabi-Yau case it sets the
right-hand side of Eq.~\eqref{dH} identically zero.

However, the low-energy gauge group is now the commutant of ${\rm SO}(6)$
within $E_8\times E_8$ which (modulo global issues) is given by
${\rm SO}(10)\times E_8$. It is interesting to compare this to the
standard Calabi-Yau case. Apparently, switching on flux has broken the
gauge group from $E_6$ to ${\rm SO}(10)$. From the decomposition
\begin{equation}
 {\bf 78}\rightarrow {\bf 45}+{\bf 16}+\overline{\bf 16}+{\bf 1}\; .
\end{equation}
of the adjoint ${\bf 78}$ of $E_6$ under ${\rm SO}(10)$ we conclude
that the additional gauge bosons in the ${\bf 16}$, $\overline{\bf 16}$
and ${\bf 1}$ representations of ${\rm SO}(10)$ must have picked
up a mass proportional to the flux parameters $e_i$, $\e_a$ and $\mu^a$.
For this to happen the additional gauge multiplets must pair up
with chiral multiplets in the same ${\rm SO}(10)$ representations.
To see how this works let us examine the decomposition of the
fundamental of $E_6$ under ${\rm SO}(10)$ which is given by
\begin{equation}
 {\bf 27}\rightarrow \bf 16+{\bf 10}+{\bf 1}\; .
\end{equation}
In the standard Calabi-Yau case, we therefore have $h^{1,1}(Y)$
chiral multiplets in $\overline{\bf 16}$ and $h^{2,1}(Y)$ chiral multiplets
in ${\bf 16}$. One ${\bf 16}$ and one $\overline{\bf 16}$ (and
one singlet) chiral multiplet have to paired up with the additional
gauge bosons, so they will pick up a mass proportional to flux
parameters. It is reasonable to expect, therefore, that
$h^{1,1}(Y)-1$ anti-families in $\overline{\bf 16}$ and $h^{2,1}(Y)-1$ families
in ${\bf 16}$ are left massless. This expectation should be
confirmed by an explicit calculation of the four-dimensional
effective theory including gauge matter. We remark
that the general formula~\eqref{Wmass} for the superpotential
should be valid including gauge matter and its evaluation
should, hence, lead to the correct gauge matter superpotential.
This will be discussed in detail in a forthcoming publication~\cite{GLM}.

\vspace{0.4cm}

A final remark concerns the gauge kinetic function $f$ of the low-energy
gauge group. From a simple reduction of the 10-dimensional gauge
field action~\eqref{SF} it is clear that, to order $\a '$,
this function is given by the dilaton, as in the standard
Calabi-Yau case. More precisely, fixing the normalization of
the gauge field kinetic term by
\begin{equation}
 -\frac{1}{4g_{\rm YM}^2}\int_{M_4}d^4x\sqrt{-g}\,{\rm Re}(f){\rm Tr}\left( F^2\right)\; ,
\end{equation}
where $F$ is the low-energy gauge field strength, and
\begin{equation}
 g_{\rm YM}^2 = \frac{4\k_{10}^2}{\a 'v}\; ,
\end{equation}
one finds that
\begin{equation}
 f = S\; .
\end{equation}
This result can be expected to receive threshold corrections at
order $(\a ')^2$ which result from terms at that order in the
10-dimensional effective action~\cite{IN}. It would be interesting
to calculate these corrections for half-flat mirror manifolds.


\section{Conclusion and outlook}

In this paper, we have considered the heterotic string on half-flat
mirror manifolds which arise in the context of mirror symmetry with
flux. More precisely, given a mirror pair $(X,Y)$ of Calabi-Yau
three-folds, the associated half-flat mirror manifolds $\hat{Y}_\z$
are the mirror duals of $X$ with NS-NS flux $\z = (e_i)$.

Our main result is the complete derivation of the four-dimensional
$N=1$ effective action to lowest order in $\a '$ on such manifolds. We
find that the K\"ahler potential for the dilaton $S$, the K\"ahler
moduli $T^i$ and the complex structure moduli $Z^a$ is the same as for
the reduction on the associated Calabi-Yau manifolds $Y$ while the
superpotential is given by
\begin{equation}
   W= \sqrt{8}\left(e_iT^i + \e_a Z^a + \m^a \cG_a\right)\; .\label{W3}
\end{equation}
Here, the first term arises from the intrinsic, geometrical flux
of the half-flat mirror manifold and the other two terms arise
from NS-NS flux with electric and magnetic parameters $\e_a$ and
$\m^a$, respectively. The structure of this result certainly
invites speculations about more general half-flat mirror manifolds
which also contain intrinsic magnetic flux and generate the
``missing'' term $m^i{\cal F}_i$ in Eq.~\eqref{W3}. Unfortunately,
at present, there is no explicit description available for such
manifolds.

We have confirmed the above result for $W$ by two largely independent methods,
namely by a reduction of the bosonic action and via a reduction
of some fermionic terms leading to the four-dimensional gravitino mass
term. As a by-product, we have also obtained a Gukov-type formula
for the superpotential which we expect to be valid for the
heterotic string on all manifolds of ${\rm SU}(3)$ structure and
includes order $\a '$ effects. It is given by
\begin{equation}
 W\sim\int_{\hat{Y}}\O \wedge \left(\hat{H}+ i dJ\right)\; ,\label{Wmass2}
\end{equation}
where $J$ is the two-form which, along with the three-form $\O$,
characterizes the ${\rm SU}(3)$ structure.

We have also argued that the standard embedding can be generalized
to the heterotic string on half-flat mirror manifolds and leads to
(in the case of $E_8\times E_8$) a low-energy gauge group
${\rm SO}(10)\times E_8$ rather than $E_6\times E_8$. We also
expect $h^{1,1}(Y)-1$ anti-families in the $\overline{\bf 16}$ representation of
${\rm SO}(10)$ and $h^{2,1}(Y)-1$ families in the
${\bf 16}$ representation.

\vspace{0.4cm}

There certainly remains substantial work to be done concerning the inclusion
of gauge and gauge matter fields. In particular, one would like to derive the
four-dimensional effective theory for these fields, understand the way
in which the flux parameters break $E_6$ to ${\rm SO}(10)$ and compute
the gauge matter superpotential explicitly. For this latter task the general
formula~\eqref{Wmass2} will be quite useful. All these issues are currently
under investigation~\cite{GLM}.

An important application of our results concerns moduli stabilization
in heterotic models. The superpotential~\eqref{W3} is independent of
$S$ so, as stands, at least the dilaton still represents a runaway
direction. However, we have seen that the gauge kinetic function
is still proportional to $S$ and, hence, gaugino condensation would
generate a non-perturbative superpotential \cite{DRSW} 
\begin{equation}
 W_{\rm gaugino} \sim \exp (-cS)
\end{equation}
for some appropriate constant $c$. Studying the combined effect of
this gaugino superpotential and \eqref{W3} is an interesting
problem which we are currently investigating.


\vspace{1cm}
\noindent
{\large\bf Acknowledgments}  S.~G.~is supported by the JSPS fellowship P03743.
A.~L.~is supported by a PPARC Advanced Fellowship. A.~M.~is supported by
PPARC. This work is supported in part by a Royal Society
Japan-UK Joint Project, reference no.~NC/Japan/JP/15207. The authors would
like to thank Mariana Gra\~na, Yukiko Konishi, Taichiro Kugo, Jan
Louis and Dan Waldram for helpful discussions.


\vskip 1cm
\appendix{\noindent\Large \bf Appendix}
\renewcommand{\theequation}{\Alph{section}.\arabic{equation}}
\setcounter{equation}{0}


\section{Some useful results on special geometry}
\label{appA}

In order to make the paper self-contained we add this appendix on
special K\"ahler geometry and their particular realizations on
Calabi-Yau manifolds. Our discussion will be carried out for a
Calabi-Yau space $Y$ with occasional reference to its mirror $X$.
 For an extensive cover of the subject see Ref.~\cite{CdO,N2}.

A K\"ahler manifold of complex dimension $n$ is called special
K\"ahler if its geometry is completely determined in terms of a
holomorphic function $\cH$, called the pre-potential. When written in
terms of projective coordinates, which we denote by $X^P$, where
$P,Q,\dots=0,\dots, n$ the pre-potential is a homogeneous function of degree
two which implies that $X^P \cH_P = 2 \cH$ with derivatives
$\cH_P = \frac{\partial \cH}{\partial X^P}$. In terms of the pre-potential,
the K\"ahler potential has the form
\begin{equation}
  \label{Ksg}
  K = - \ln{i \left(\bar X^P \cH_P - X^p \bar \cH_P \right)}\; .
\end{equation}
It is also useful to introduce a $(n+1) \times (n+1)$ matrix $\cQ$
\begin{equation}
  \label{Nsg}
  \cQ_{PQ} = \bar \cH_{PQ} + 2i \frac{\mathrm{Im}( \cH_{PR}) \mathrm{Im}
  (\cH_{QS}) X^Q X^S}{\mathrm{Im}( \cH_{RS})X^R X^S} \; ,
\end{equation}
which plays the role of gauge coupling matrix in type II
compactifications and which satisfies $\cH_P = \cQ_{PQ} X^Q$.

It is well-known that the moduli space of Calabi-Yau
manifolds is governed by two such special K\"ahler geometries: one for
the complexified K\"ahler moduli and one for the complex structure
moduli. Let us now describe these two moduli spaces in turn.

\vspace{.4cm}

We start with the K\"ahler moduli space of the Calabi-Yau manifold $Y$
which has dimension $n=h^{1,1}(Y)$. We denote its projective coordinates
by $T^I$ with indices $I,J,\dots = 0,\dots ,h^{1,1}(Y)$. It is also
useful to introduce indices $i,j,\dots = 1,\dots ,h^{1,1}(Y)$.

In the large radius limit of the Calabi-Yau space, the pre-potential, which
we call ${\cal F}$, is known explicitly and given by
\begin{equation}
  \label{Kprep}
  \cF = -\frac{1}{6}\frac{d^{(Y)}_{ijk} T^i T^j T^k}{T^0}\; ,
\end{equation}
where $d^{(Y)}_{ijk}$ are the triple intersection numbers of the manifold $Y$.
Introducing affine coordinates $t^i=T^i/T^0$, one finds from Eq.~\eqref{Ksg}
for the associated K\"ahler potential
\begin{equation}
  \label{KKs}
  \KK = - \ln\left( i d_{ijk}^Y (t^i-\bar t^i)(t^j - \bar t^j) (t^k -
  \bar t^k)\right) \equiv - \ln{ 8 \V} \; ,
\end{equation}
where ${\cal V}$ can be interpreted as the volume of the Calabi-Yau space.
It is useful to describe the moduli space in terms of the K\"ahler
form $J$ which can be expanded as
\begin{equation}
 J = v^i\o_i\; ,
\end{equation}
where $v^i=\mathrm{Im}(t^i)$ and $(\o_i)$ is a basis of the second
cohomology of $Y$. Then, the metric  $\gK_{ij}$ on the K\"ahler
moduli space can be written as
\begin{equation}
 \gK_{ij}=\partial_i\bar{\partial}_j\KK=\frac{1}{4{\cal V}}\int_{Y}\o_i\wedge
          \star\o_j\; .
\end{equation}
A useful relation which can be derived from the explicit K\"ahler
potential~\eqref{KKs} is
\begin{equation}
  \label{uf}
  \gKi^{ij} \KK_j = (t^i - \bar t^i) = 2i v^i \; ,
\end{equation}
where $\KK_i$ denote the derivatives of the K\"ahler
potential~\eqref{KKs} with respect to the fields $t^i$ and $\gKi^{ij}$
is the inverse of the K\"ahler metric.

One can also explicitly compute the coupling matrix defined in \eqref{Nsg}
which we denote by $\cN$. The components of $\cN$ together
with the ones of $(\mathrm{Im}(\cN ))^{-1}$ are given by
\begin{equation}
  \label{eq:N}
  \begin{aligned}
    \mathrm{Re}( \cN_{ij}) & = - d^{(Y)}_{ijk} b^k \ , &
    \mathrm{Im}( \cN_{ij}) & = -4 \V \gK_{ij}\; , \quad  &
    (\mathrm{Im}( \cN))^{-1 \; ij} & = - \frac{\gKi^{ij}}{4 \V} -
    \frac{b^i b^j}{\V}\; . \\
    \mathrm{Re}( \cN_{i0}) & = \frac12 d^{(Y)}_{ijk} b^j b^k \ , &
    \mathrm{Im}( \cN_{i0}) & =  4\V \gK_{ij} b^j \: , &
    (\mathrm{Im}( \cN))^{-1 \; i0} & = - \frac{b^i}{\V} \\
    \mathrm{Re} (\cN_{00}) & = - \frac13 d^{(Y)}_{ijk} b^i b^j b^k \, ,
    \quad &  \mathrm{Im} (\cN_{00}) & = -  \V - 4 \V \gK_{ij} b^i b^j \;
    , & \quad (\mathrm{Im}( \cN ))^{-1 \; 00} & = - \frac1\V\; .
  \end{aligned}
\end{equation}

\vspace{.4cm}

Let us now pass to the complex structure moduli space of the same
Calabi-Yau manifold $Y$ which has dimension $n=h^{2,1}(Y)$.  We denote
the projective coordinates on this moduli space by $Z^A$, where
$A,B,\dots =0,\dots ,h^{2,1}(Y)$ and also introduce lower-case indices
$a,b,\dots = 1,\dots ,h^{2,1}(Y)$. The pre-potential is called ${\cal
G}$. In general, an explicit expression for this pre-potential cannot
be written down. However, one can still derive some useful formulae
when working with a generic $\cG$. Most of the properties of this
space can be described in terms of the holomorphic $(3,0)$ form
$\Ox$. Recall that in a real, symplectic basis $(\a_A,\b^B)$ of
three-forms it can be expanded as
\begin{equation}
  \label{Oexp}
  \Ox = Z^A \ax_A  - \cG_A \bx^A \; .
\end{equation}
It follows immediately that the K\"ahler potential can be written as
\begin{equation}
  \label{Kcs}
  \Kcs =  - \ln\left( i ( \bar Z^A \cG_A - Z^A \bar \cG_A) \right)
       =- \ln\left( i \int_{Y} \Ox \wg \bar \Ox\right)\; .
\end{equation}
Let us here denote the coupling matrix~\eqref{Nsg} by $\cM$. It
turns out that, for the complex structure moduli space, this matrix
has a proper geometric interpretation in terms of the integrals
\begin{eqnarray}
  \label{ABC}
  B_{AB} = \int_{Y_3} \ax_A \wg *\ax_B & = & \int_{Y_3} \ax_B \wg *\ax_A
  = B_{BA}\ , \nn  \\
  C^{AB} = - \int_{Y_3} \bx^A \wg *\bx^B & = & - \int_{Y_3} \bx^B \wg *
  \bx^A = C^{BA}\ ,  \\
  {A_A}^B = - \int_{Y_3} \bx^B \wg *\ax_A & = & - \int_{Y_3} \ax_A
  \wg * \bx^B   \; , \nn
\end{eqnarray}
which can be expressed as~\cite{HS,CDF}
\begin{eqnarray}
  \label{A-N}
  A & = & \mathrm{Re}( \cM ) \left(\mathrm{Im}( \cM )
  \right)^{-1}\ , \nn \\
  B & = & - \mathrm{Im}( \cM ) - \mathrm{Re}( \cM )
   \left(\mathrm{Im} (\cM )\right)^{-1} \mathrm{Re}( \cM )\ ,  \\
  C & = & \left(\mathrm{Im} (\cM ) \right)^{-1}\ . \nn
\end{eqnarray}

A particularly useful insight can be obtained by choosing a
different basis for the third cohomology of $Y$. One can define
complex $(2,1)$ forms $\chi_a$ via Kodaira's formula~\cite{KOD}
\begin{equation}
  \frac{\partial\O}{\partial z^a} = -\Kcs_a\O + \chi_a
\end{equation}
where $z^a=Z^a/Z^0$ are the affine coordinates and $\Kcs_a$ denote
the derivatives of the complex structure K\"ahler potential \eqref{Kcs}
with respect to $z^a$. Then the forms $(\Ox
,\chi_a,\bar\chi_a,\bar\Ox)$ form a basis for the third cohomology of
$Y$. In this new basis, the metric $\gcs_{ab}$ on the complex structure moduli
space has the simple form
\begin{equation}
  \label{gcs}
    \gcs_{a\bar{b}} \equiv \partial_a \partial_{\bar b} \Kcs=
  -\frac{\int_{\hat{Y}} \c_a \wedge \bar{\c}_{\bar{b}}}{\int_{\hat{Y}}
  \O \wedge \bar{\O}} \; .
\end{equation}
The transformation from the symplectic basis $(\ax_A ,\bx^A)$ to the
complex basis defined above can be summarized as
\begin{equation}
  \label{abeta}
  \begin{aligned}
    \b^A = & \tilde f^{A}\O + \tilde f^{Aa}\chi_a + \mbox{h.c.}\; ,\\[2mm]
    \a^A = & f_{A}\O + f_{A}{}^a \chi_a + \mbox{h.c.}\; ,
  \end{aligned}
\end{equation}
where
\begin{equation}
  \label{fh}
  \begin{aligned}
    \tilde f^A & = -\frac{\bar Z^A}{\int_Y \O \wedge \bar \O} \; , \qquad &
    \tilde f^{Aa} & = \frac{\gcsi^{a \bar b}}{\int_Y \O \wg \bar\O} \bar
    D_{\bar b} \bar Z^A \; , \\[3mm]
    f_A & = -\frac{\bar \cG_A}{\int_Y \O \wg \bar \O} \; ,&
    f_A{}^a & = \frac{\gcsi^{a\bar b}}{\int_Y \O \wg \bar \O}
    \bar D_{\bar b} \bar \cG_A \; ,
  \end{aligned}
\end{equation}
and by $\gcsi^{a \bar b}$ we denote the inverse of the metric
\eqref{gcs}. The K\"ahler covariant derivatives $D$ are defined by
\begin{equation}
  \begin{aligned}
    \bar D_{\bar b} \bar Z^A & =  \partial_{\bar b}\bar Z^A
    +\overline{\Kcs_{\bar b}}\bar Z^A  \; ,\\
    \bar D_{\bar b} \bar \cG_A & = \partial_{\bar b}\bar \cG_A
    +\overline{\Kcs_{\bar b}}\bar \cG_A \; .
  \end{aligned}
\end{equation}

Until now all the formulae for the complex structure moduli space were
generic and can be applied to any Calabi-Yau manifold.  However, in
the limit of \emph{large complex structures} one can be somewhat more
explicit. For this we rely on mirror symmetry which relates the
complex structure deformations of the Calabi-Yau manifold $Y$ to
K\"ahler deformations on the mirror $X$. As a result, the
pre-potential $\cG$ is now given by a cubic formula similar to
Eq.~\eqref{Kprep}, that is,
\begin{equation}
 \label{csprep}
  \cG = - \frac16 \; \frac{d^{(X)}_{abc} Z^a Z^b Z^c}{Z^0} \; .
\end{equation}
Here, $d^{(X)}_{abc}$ are the triple intersection numbers of the
mirror Calabi-Yau manifold $X$. The matrix $\cM$ can be computed explicitly
in this limit and is given by
\begin{eqnarray}
  \label{eq:M}
  \mathrm{Re} (\cM_{00}) & =& - \frac{1}{24} d^{(X)}_{abc} (Z^a + \bar
    Z^a) (Z^b + \bar Z^b) (Z^c + \bar Z^c)\, , \nn \\
    \mathrm{Im} (\cM_{00}) &=& -\frac{e^{-\Kcs}}{8} [1+
    \gcs_{ab} (Z^a + \bar Z^a) (Z^b + \bar Z^b)]\, ,  \nn \\
    \mathrm{Re} (\cM_{a0}) &=& \frac18 d^{(X)}_{abc} (Z^b + \bar Z^b)
    (Z^c + \bar Z^c) \, ,\nn \\
    \mathrm{Im} (\cM_{a0}) &=&  \frac{e^{-\Kcs}}{4} \gcs_{ab} (Z^b +
    \bar Z^b) \, , \\
    \mathrm{Re} (\cM_{ab}) &=& - \frac12 d^{(X)}_{abc} (Z^c + \bar Z^c) \ , \nn\\
    \mathrm{Im} (\cM_{ab}) &=& - \frac{e^{-\Kcs}}{2} \gcs_{ab}\, . \nn
\end{eqnarray}
The components of $(\mathrm{Im}(\cM ))^{-1}$ read
\begin{eqnarray}
  \label{M-1}
  (\mathrm{Im} (\cM ))^{-1 \; ab} & = & - 2 e^{\Kcs} \Big[\gcsi^{ab} +
  (Z^a + \bar Z^a) (Z^b + \bar Z^b) \Big] \; , \nn \\
  (\mathrm{Im} (\cM ))^{-1 \; a0} & = & - 4  e^{\Kcs} (Z^a + \bar Z^a)
  \; ,\\
  (\mathrm{Im} (\cM ))^{-1 \; 00} & = & - 8 e^{\Kcs} \; . \nn
\end{eqnarray}

\vspace{.4cm}

As a simple application of the above formulae and as a warm-up for
the next section we can rewrite the potential~\cite{TV},
obtained by turning on $H$--fluxes in Calabi-Yau
compactifications, in a more suggestive way which makes
it easier to read off the superpotential. As in Eq.~\eqref{Hflux},
the $H$--flux~\footnote{Unlike in the main part of the paper, $\e_A$ and
$\mu^A$ denote arbitrary flux parameters, that is, we allow $\e_0\neq 0$ and
$\m^0\neq 0$.}
\begin{equation}
  H = \e_A \b^A + \m^A \a_A \; .
\end{equation}
can be expanded in terms of the symplectic basis $(\a_A,\b^B)$.
With Eqs.~\eqref{ABC}, this potential can be written as
\begin{equation}
  \label{VH}
  e^{-K}V_H = 4e^{-K^{(2)}}\int H\wedge *H = -4e^{-K^{(2)}}(\e_A
  +\mu^C \cM_{AC} ) \mathrm{Im} \cM^{-1AB}(\e_B + \mu^D \bar\cM_{BD}).
\end{equation}
On the other hand, writing the $H$--flux in the complex basis defined
in \eqref{abeta} the above formula reads
\begin{eqnarray}
  e^{-K} V_H & = & 8 e^{-K^{(2)}}(\e_A \tilde f^{Aa} + \m^A f_A{}^a)(\e_B
  \bar{\tilde f}^{Bb} + \m^B \bar f_B{}^b) \int \chi_a \wg * \bar \chi_b \nn\\
  & & + 8e^{-K^{(2)}}|\e_A \tilde f^A + \m^A f_A|^2 \int \O \wg *\bar \O.
\end{eqnarray}
Inserting the relations \eqref{fh} and using \eqref{gcs} we obtain
\begin{eqnarray}
  e^{-K}V_H & = & 8 \gcsi^{ab}(\e_A D_a Z^A + \m^A D_a \cG_A)(\e_B\bar
  D_b \bar Z^B + \m^B \bar D_b \bar \cG_B) \nn \\
  && + 8|\e_A Z^A + \m^A \cG_A|^2 \;
\end{eqnarray}
Thus we can write
\begin{equation}
  \label{VWH}
  V_H =  \gcsi^{a \bar b}  (D_a W_H) (\overline{D_b W_H}) + |W_H|^2 \; ,
\end{equation}
where we have defined
\begin{equation}
  \label{WH}
  W_H = \sqrt 8 \left( \e_A Z^A + \m^A \cG_A \right) \; .
\end{equation}
Let us stress that, at this level, Eq.~\eqref{VWH} does not yet have
the structure of the usual supergravity relation~\eqref{VSUGRA}
between the potential and the superpotential since the term $-3|W|^2$
is not correctly reproduced. However, for heterotic strings
compactified on Calabi-Yau manifolds in the presence of $H$--fluxes
also the dilaton and the K\"ahler moduli have to be included in
calculating the potential. Their contribution is precisely $4 |W_H|^2$
which cancels against $-3 |W|^2$ leaving behind precisely the factor
$|W_H|^2$ present in \eqref{VWH}.


\section{Superpotential including NS-NS flux}
\label{appB}

Having defined all the technical tools in the previous section, we
are now ready to show that the scalar potential in Eq.~\eqref{Vflux}
can be indeed obtained from the superpotential~\eqref{W1} using
the general supergravity formula~\eqref{VSUGRA}. To do this it
will be useful to replace  $\tilde{\e}_0 = e_i b^i$ in the
potential~\eqref{Vflux} and pull apart the contributions to the
potential coming from the torsion of the half-flat mirror manifold
and the one coming from the $H$--flux, writing the potential as
\begin{eqnarray}
  \label{fullred}
  V = V_T + V_H + V_{\mathrm{mix}}\; .
\end{eqnarray}
Here, $V_{T}$ arises from the torsion of the internal manifold,
$V_{H}$ is due to $H$--flux and $V_{\mathrm{mix}}$ is the mixed
term which is present when both are taken into account
simultaneously. Explicitly, these parts are given by
\begin{eqnarray}
  \label{expV}
  V_{T} & = & -2e^{2\phi + K^{(2)}} e_i\left(\mathrm{Im}(\cN)\right)^{-1ij} e_j \; ,
  \nn \\
  V_{H} & = & -2e^{2\phi + K^{(1)}}(\e_A + \mu^C \cM_{AC} )
  \left(\mathrm{Im}(\cM )\right)^{-1AB}(\e_B + \mu^D \bar\cM_{BD}) \; ,\\
  V_{\mathrm{mix}} & = & - 4(e_i b^i) e^{2\phi + K^{(1)}} \Big[ \e_a
  \left(\mathrm{Im}(\cM )\right)^{-1\, 0a} + \m^c \left(\mathrm{Im}(\cM )
  \right)^{-1 \, 0A} \mathrm{Re}
  \cM_{Ac} \Big] \; ,\nn
\end{eqnarray}
where, in the second line, we have used the convention
\begin{equation}
 \e_A  = (0,\e_a)\; ,\qquad \m_A  = (0,\m_a)\; .
 \label{em0}
\end{equation}
Taking into account Eq.~\eqref{eq:N} one finds that $V_T$ defined
above is precisely the potential obtained in Eq.~\eqref{V} while $V_H$
is the potential we have discussed in Eq.~\eqref{VH}.

Let us split the superpotential in its two main pieces
\begin{equation}
  \label{WTH}
  W= W_T + W_H \; ,
\end{equation}
where $W_T$ was defined in Eq.~\eqref{W} and $W_H$ is taken from
Eq.~\eqref{WH} with the specific flux parameters~\eqref{em0} inserted.
Note that $W_T$ depends only on the K\"ahler moduli $T^i$ while $W_H$
depends only on the complex structure moduli $Z^a$. We would like to
reproduce the potential~\eqref{fullred} by inserting this superpotential,
as well as the standard K\"ahler potential~\eqref{K}, into the general
supergravity formula~\eqref{VSUGRA}.We start by evaluating the K\"ahler
covariant derivatives which now read
\begin{equation}
  \label{KcdW}
  \begin{aligned}
    D_S W & =  -\frac12 e^\phi \, W \; ,\\
    D_i W & = \sqrt 8 e_i + \KK_i W \; ,\\
    D_a W & = D_a W_H + \Kcs_a W_T \; .
  \end{aligned}
\end{equation}
Using the K\"ahler metric $\gK_{ij}$ in terms of the fields $t^i$,
to be derived from the K\"ahler potential~\eqref{KKs}, we obtain
\begin{eqnarray}
  D_S W \, (\overline{D_S W}) K^{\bar S S} & = & |W|^2 \nn \\
  D_i W \, (\overline{D_j W}) K^{\bar j i} & = & 8 e_i e_j \gKi^{ij} -
  2 i \sqrt 8 (e_i v^i)(W - \bar W) + 3 |W|^2 \\
  D_a W \, (\overline{D_b W}) K^{\bar b a} & = & D_a W_H
  (\overline{D_b W_H}) \gcsi^{\bar b a} + (\Kcs_a \overline{\Kcs_b}
  \gcsi^{\bar b a})|W_T|^2\nn\\
  &&+ \left[ D_a W_H (\overline{\Kcs_b W_T}) \gcsi^{\bar b a} +
  \mathrm{h.c.} \right]\; , \nn
\end{eqnarray}
where we have used
\begin{equation}
  \KK_i\overline{\KK_j} \gKi^{ij} = 3 \; ,
\end{equation}
which follows for the cubic pre-potential~\eqref{Kprep}.

In the large complex structure limit, the pre-potential for the complex
structure moduli space is given by \eqref{csprep} and thus, in analogy
with the K\"ahler moduli space, we have $\Kcs_a (\overline{\Kcs_b})
\gcsi^{\bar b a} = 3$. Using this relation and further splitting $W$
as in Eq.~\eqref{WTH} we obtain from the supergravity formula
\eqref{VSUGRA}
\begin{eqnarray}
  \label{VWint}
  e^{-K} V & = & 8 e_i e_j \gKi^{ij} - 32 (e_i v^i)^2 +
  4|W_T|^2 \nn \\
  & & + D_a W_H (\overline{D_b W_H}) \gcsi^{\bar b a} + |W_H|^2 \\
  & & + (W_T - \bar W_T)(W_H - \bar W_H) + 4 (W_T \bar W_H + \bar W_T
  W_H) + \left[ \partial_a W_H (\overline{\Kcs_b}) \gcsi^{\bar b a}
  \bar W_T + \mathrm{h.c.} \right]\, . \nn
\end{eqnarray}
It was shown in the main part of the paper that the first line in the
above equation reproduces the potential $V_T$ from Eq.~\eqref{expV} while
we have proved in Eq.~\eqref{VWH} that the second line gives rise to the
potential $V_H$.

In order to evaluate the mixed terms in the last line which contain
half-flat as well as $H$--flux, we need to compute the expressions
$\partial_a W_H (\overline{\Kcs_b})\gcsi^{\bar b a}$. In general
this is a complicated task, but in our case, as we work in the large
complex structure limit, this computation is fairly easy. First of all
note that in this case we can derive a formula similar to Eq.~\eqref{uf},
namely
\begin{equation}
  \gcsi^{a \bar b} (\overline{\Kcs_b}) = - (z^a - \bar z^a) \; .
\end{equation}
Then, making explicit use of the cubic formula for the
pre-potential $\cG$, Eq.~\eqref{csprep}, one can rewrite the
potential~\eqref{VWint} into the following form
\begin{eqnarray}
  \label{VWfin}
  V & = & V_T + V_H + 32 (e_i b^i) e^K  \e_a(Z^a+\bar Z^a) \\
  & & + 8e^K \left[e_i \bar T^i \m^a (\bar \cG_a + 3 \cG_a + (\bar Z^b -
  Z^b)\cG_{ba}) + h.c. \right]  \; .\nn
\end{eqnarray}
Working from the other end, we now rewrite the mixed part of the potential
$V_{\mathrm{mix}}$ in Eq.~\eqref{expV}, obtained from the reduction, by
using the explicit form of the matrix $\cM$ given in Eq.~\eqref{eq:M}.
One can easily show that
\begin{equation}
  \left(\mathrm{Im}(\cM )\right)^{-1 \, 0A} \mathrm{Re}(\cM )_{Ac} = -2 e^{\Kcs}
  (\bar \cG_a + 3 \cG_a + (\bar Z^b -  Z^b)\cG_{ba}) \; .
\end{equation}
Inserting this relation into Eq.~\eqref{expV} and using the
expression for $(\mathrm{Im}(\cM ))^{-1 \, 0a}$ from \eqref{M-1}
one obtains
\begin{equation}
  V_{\mathrm{mix}} = 32 (e_i b^i) e^K  \e_a(Z^a+\bar Z^a)
  + 8e^K \left[e_i \bar T^i \m^a (\bar \cG_a + 3 \cG_a + (\bar Z^b -
  Z^b)\cG_{ba}) + h.c. \right] \; .
\end{equation}
Comparing the result \eqref{VWfin} from the supergravity side with
the reduction result \eqref{fullred}, where $V_{\mathrm{mix}}$
is given by the formula above, we see that the two potentials are
indeed the same. This proves that the potential obtained by
compactifying the heterotic string on half-flat mirror manifolds with
$H$--flux \eqref{Vflux}, can be obtained from the $N=1$
supergravity formula \eqref{VSUGRA} with the superpotential given by
\eqref{W1}.


\end{document}